# High-power photoconductive THz emitters for fast parallel electro-optic detection

Mohsen Khalili,[1,*] Robin Löscher[1], Anna Martinez[1], Abhishek Singh[2], Stephan Winnerl,[3] and Clara J. Saraceno[1,4]

[1]Photonics and Ultrafast Laser Science, Ruhr-University Bochum, 44801, Germany
[2]Centre for Advanced Electronics, Indian Institute of Technology Indore, 453552, India
[3]Helmholtz-Zentrum Dresden-Rossendorf, Institute of Ion Beam Physics and Materials Research, 01328 Dresden, Germany
[4]Research Center Chemical Science and Sustainability, University Alliance Ruhr, 44801 Bochum, Germany

*mohsen.khalilikelaki@ruhr-uni-bochum.de

The terahertz (THz) region is highly attractive for imaging because it penetrates many optically opaque materials and can distinguish them by composition. Among the various methods that allow us to access this spectral region, THz time-domain spectroscopy (THz-TDS) is uniquely suited to retrieve broadband information, recording the full waveform of the THz electric field to recover both amplitude and phase with high sensitivity. Translating this capability into fast, wide-field imaging has however remained a major challenge, largely because the low average power of conventional TDS sources forces a trade-off between acquisition speed, field of view, and sensitivity. Here we demonstrate a hyperspectral THz camera in which a high-average-power, pulsed broadband THz source, delivering 38.3 mW at a 400 kHz repetition rate, drives parallel electro-optic detection over a wide field of view at high acquisition rates. The system simultaneously records the full THz waveform, in amplitude and phase, across 145,000 pixels over a ~14.5 mm² field of view at an acquisition rate of 2 Hz. These results demonstrate the potential of high-average-power THz-TDS for broadband, high-speed imaging, opening applications such as inline material-resolved inspection and the spectroscopic imaging of irreversible, second-timescale processes such as hydration in biologically relevant systems.

## Introduction

The THz band occupies a distinctive position in the electromagnetic spectrum: its photon energies coincide with a wealth of low-energy phenomena such as lattice phonons, intermolecular and hydrogen-bond modes, and free-carrier transport in semiconductors[1–5]. Terahertz time-domain spectroscopy (THz-TDS) probes these directly with high sensitivity by measuring the full waveform of the THz field, retrieving its amplitude and phase. This makes THz-TDS a versatile method, used across many scientific disciplines including condensed-matter physics and chemistry as well as industrial inspection for non-destructive testing, as it gives access to the full dielectric function of samples over a broad spectral range. Extending these capabilities from spatially averaged spectroscopy to imaging is an active topic of research and many efforts are currently dedicated to spatially resolving the THz field at sufficient speed.

The conventional path to achieve spatial resolution is raster-scanning the sample point by point and using a single detector, including near-field schemes to reach sub-diffraction resolution[6,7]. This scheme makes it possible to recover the full spectroscopic content at every point but is inherently serial, with typical acquisition times from minutes to hours. This throughput limit keeps THz-TDS away from practical imaging applications: large or heterogeneous samples, statistically meaningful fields of view, and time-evolving processes are all out of reach at these slow rates. Generally, high-throughput spectrally resolved THz imaging remains a missing cornerstone that would unblock many fields, including real-world applications in fast inspection, but also the observation of many irreversible phenomena in fundamental research, for example mapping dielectric properties of water in cells or other hydration changes in biologically relevant systems[4,8,9].

This throughput bottleneck has motivated a range of faster strategies, from high-speed scanning to parallel-detector and computational imaging[10] (a detailed table of the different state-of-the-art systems presented below is given in the supplementary material, Table S1). High-speed asynchronous-optical-sampling systems reach pixel rates of order 10 kHz while preserving amplitude and phase[11] but still build images sequentially. Computational schemes based on structured illumination patterns and single-pixel detection in combination with compressed sensing reduce the number of measurements needed to form an image even with deeply sub-wavelength resolution through near-field probing[12,13] but rely on iterative reconstruction and recover only limited spectral or temporal information per pixel.

Parallel-detector schemes circumvent the difficulties of raster scanning by capturing the entire field-of-view in one measurement. Two implementations stand out in this regard: photoconductive arrays and electro-optic imaging.

Plasmonic photoconductive focal-plane arrays were recently demonstrated with a full broadband implementation (up to 3 THz) achieving 16 fps frame rate with ~250 µm spatial resolution and even super-resolution down to ~60 µm by deep-learning-based reconstruction[14]. This is a very promising implementation; however, it requires dedicated fabrication, remaining a highly specialized device with much of the performance fixed during fabrication.

Electro-optic (EO) imaging offers another highly attractive alternative. Here, the THz field is mapped via the electro-optic effect onto an optical probe beam when impinging on a nonlinear crystal with $\chi^{(2)}$ nonlinearity. Rather than using a single-pixel THz detector as in usual THz-TDS, a CCD camera records spatially resolved probe images. First demonstrated 30 years ago[15], it was later extended to near-field implementations with resolutions down to 14 µm (λ/30 at 0.7 THz) [16], and even subsequently extended to λ/600 (~5 µm at 100 GHz) using thin detection crystals[17] and megapixel detection at far field imaging[18]. The main element of imaging in this technique is a widely commercially available bulk nonlinear crystal typically offering large parameter flexibility in terms of bandwidth, sensitivity and aperture. Despite great promise, what has held EO imaging back is the low average power of THz sources: each pixel requires a minimum THz field for EO sampling, therefore prior demonstrations could only be performed using kHz repetition rate laser systems, making short acquisition time-domain detection impossible to realize and restricting this implementation to a few academic uses, mostly due to the use of high peak power, scientific laser driving systems.

In this context, overcoming the average-power bottleneck in THz-TDS is an active field of research, and several concepts driven by modern high-power ultrafast Ytterbium lasers now make it routinely possible to reach tens to hundreds of milliwatts of THz average power, with record systems approaching the watt level[19]. Optical rectification in inorganic crystals is the most commonly used route, with the tilted-pulse-front geometry in lithium niobate delivering 66 mW at MHz repetition rates[20] and, more recently, 643 mW at 40 kHz[21]. Organic crystals offer higher nonlinear coefficients and broader bandwidth and have recently reached the ten-milliwatt level in MNA at 100 kHz[22]. Two-color plasma sources provide the broadest spectra and have been scaled to 640 mW at 500 kHz[23]. Photoconductive devices, finally, are attractive because their pulse-energy requirements are more modest than those requiring high peak powers, offering a route to high average powers from simpler driving systems. Very recent results using large-area photoconductive emitters demonstrate multi-milliwatt average powers at hundreds of kHz[24]. However, so far, the potential of none of these advances in high-power sources for fast parallel broadband imaging has not been shown.

Here we demonstrate wide-area, spatially and spectrally-resolved EO detection at high speed, thanks to a record-high average power photoconductive large-area THz emitter (LAE). The source is driven by a commercial, industrially robust Yb-based femtosecond laser and achieves a THz average power of 38.3 mW, which is to the best of our knowledge the highest reported to date from any pulsed photoconductive THz emitter. Employing this source in a parallel EO imaging geometry, we reconstruct pixel-resolved THz waveforms with a ~1.5 THz bandwidth across a 14.5 mm² circular field of view (4.3 mm equivalent diameter, ~145,000 pixels at 10 µm pixel size) with a single-trace acquisition time of 500 ms. This corresponds to an acquisition throughput of ~$2.9 \times 10^5$ sampled pixel waveforms per second at the camera's 10 µm sampling pitch. The fastest previously reported fully spectrally resolved THz imaging system reached 23,200 px/s, quoted at diffraction-limited resolution and for one-dimensional scanning[25]; because that metric is resolution-normalized while ours is based on the camera sampling grid rather than the diffraction-limited resolution of our current setup (~0.3–1.3 mm, see Discussion), the comparison reflects the gain in acquisition throughput rather than a directly comparable, resolution-normalized benchmark. The current design could be adapted to a variety of application needs, including wider fields of view, sub-diffraction resolution, higher acquisition speed, and higher bandwidth.

## Results

### 1. Record high average power photoconductive large area emitter

#### Laser system and emitter configuration

The experimental setup including the source and its integration into a standard spatially integrated THz-TDS is presented in **Fig. 1**a. The THz source is based on a semi-insulating GaAs (SI-GaAs) microstructured large-area photoconductive emitter with an interdigitated electrode metal-semiconductor-metal (MSM) structure, described in refs [26,27]. The emitter has an active area of 10×10 mm². The emitter is housed in a water-cooled copper mount (**Fig. 1**b) surrounding the LAE with a 10-mm-diameter clear aperture and providing heat removal through the emitter edges. **Fig. 1**c shows the geometry of the LAE. The MSM structure is masked by a second metallization layer that is electrically isolated from the MSM electrodes and blocks optical excitation in every second gap. In this way, charge carriers are excited and accelerated unidirectionally and radiated THz fields coherently combine in the far field. The electrode structure, with a width and spacing of 10 µm, can produce a bias field on the order of tens of kV/cm at a moderate voltage in the range of tens of V.

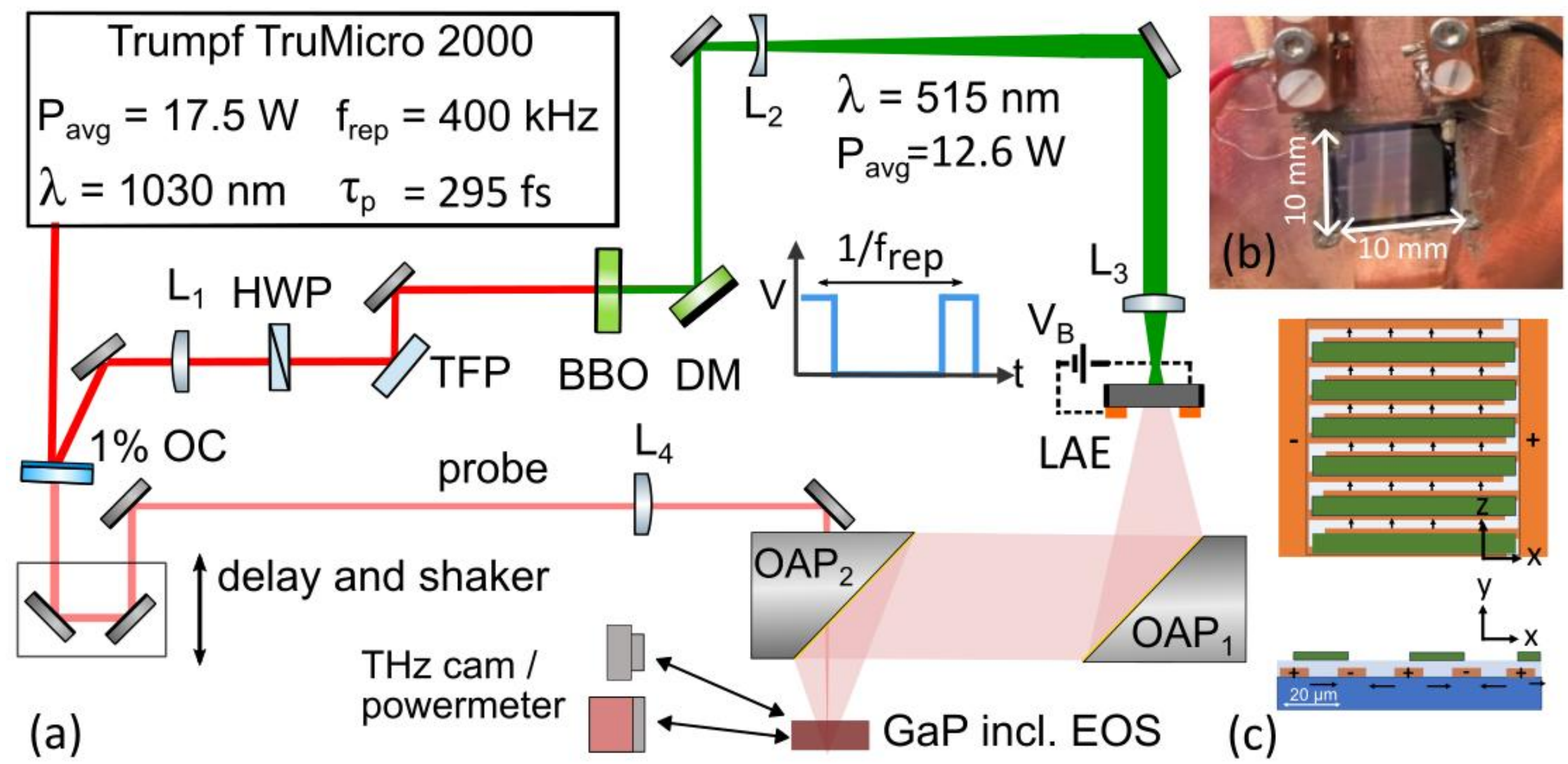

**Fig. 1 | Experimental setup and photoconductive emitter. a** Experimental setup for THz generation and detection in a standard TDS configuration. HWP, half-wave plate; TFP, thin-film polarizer; DM, dichroic mirror. **b** LAE in the water-cooled copper mount. **c** Geometry of the microstructured large area emitter (LAE). The pump is focused onto the emitter to a spot diameter of 7 mm ($1/e^2$ intensity), the emitter being positioned after the focal plane of the pump lens at the point of maximum collected THz power. The emitted THz radiation is collected and refocused onto the detection crystal by a pair of 90° off-axis parabolic mirrors with reflected focal lengths of 101.6 mm and 50.8 mm, separated by the sum of their focal lengths.

The LAE is excited by a commercially available Yb-based laser amplifier delivering up to 17.5 W of average power at a central wavelength of 1030 nm with a pulse duration of 295 fs and a maximum repetition rate of 400 kHz (TruMicro2000, TRUMPF). This laser system is originally designed for material processing applications, and is thus a turn-key, industrially robust fs-laser. For excitation of charge carriers in GaAs (bandgap energy ~1.42 eV), the laser is frequency-doubled in a 500-µm-thick BBO crystal, yielding up to 12.6 W of average power at 515 nm (photon energy 2.41 eV), with a Gaussian-equivalent pulse duration of approximately 253 fs obtained from the intensity-autocorrelation FWHM using a Gaussian deconvolution. The LAE is positioned after the focal plane of the pump lens, such that the pump spot size on the emitter can be adjusted by translating the device along the beam-propagation direction. Apart from small diffraction-related deviations affecting the longer-wavelength spectral components, the generated THz beam preserves the phase front of the optical pump beam[24,28]. For EOS measurements, a 1% output coupler directs a small fraction of the laser power into the probe arm, where a fast-oscillating delay line (3 Hz, 15 ps scan range) periodically delays the probe pulse with respect to the THz pulse. The optical spectra and intensity autocorrelations of the 1030 nm fundamental and its 515 nm second harmonic are presented in Supplementary Fig. S7.

**THz source and TDS performance**

The detailed optimization procedure for the emitter is presented in the Methods. The device is operated using synchronized pulse-bias operation, which allows us to apply a large bias field up to 95 kV/cm, which is the main driver of THz generation efficiency in these photoconductive devices. The optimal conditions were reached using a unipolar pulse-width modulation (PWM) bias with 650 ns voltage pulses, corresponding to a duty cycle of 26%, synchronized to the 400 kHz laser repetition rate. Under these conditions, a record-high THz average power of 38.3 ± 0.2 mW was obtained at 12.6 W of optical pump power, corresponding to an optical-to-THz conversion efficiency of 0.30%. To our knowledge, this is the highest average power reported to date from any pulsed photoconductive THz emitter. **Fig. 2**a shows the THz electric field measured by electro-optic sampling (EOS) under high-power operation, using a 3-mm-thick GaP crystal. For this measurement, the optical pump power was 12.6 W with the previously determined optimized pump spot diameter of 7 mm, and the bias voltage was set to 84 V, equivalent to 84 kV/cm of acceleration field (the acceleration field is quoted as the applied voltage divided by the 10 μm electrode spacing). This measurement was therefore not taken at the record operating point. The bias was held at 84 kV/cm rather than at the 95 kV/cm used for the power record, as a conservative margin against breakdown over the long acquisition; at this bias the emitter delivers 34.6 ± 0.2 mW. The normalized waveform is already close to its high-excitation limit at this point, since both the pump power and the acceleration field lie above the values beyond which the normalized waveform no longer changes, as shown in Supplementary Note 3. The bias modulation was synchronized at half the laser repetition rate (200 kHz) for low-noise lock-in detection. A reference dark trace was recorded by switching off the bias voltage while keeping all other experimental conditions unchanged.

The corresponding spectra and dynamic range (DR) characterization are shown in **Fig. 2**b–d. With the 253 fs excitation pulses used for the main high-power measurements, the spectrum extends to approximately 3 THz and reaches a peak DR of 102 dB for a total acquisition time of 156 s (935 full waveforms). **Fig. 2**c shows the evolution of the peak DR as a function of the number of averaged traces, estimated using a random-subset averaging procedure in which a defined number of traces is randomly selected without replacement, averaged in the time domain, and the spectral peak DR extracted; this procedure is repeated multiple times to obtain the mean and standard deviation. The DR scales reproducibly with averaging, confirming that the high DR results from the accumulation of stable, reproducible waveforms rather than from occasional exceptional traces. This is further confirmed by **Fig. 2**d, where individual THz waveforms are plotted as a function of trace index: the main THz transient remains at a fixed temporal position with no observable timing drift or systematic amplitude degradation throughout the acquisition.

The spectrum of our source peaks at low frequencies and extends to approximately 3 THz. To test whether the thickness-dependent response of the 3-mm GaP crystal limits the measured spectral extent, we repeated the measurement under identical excitation conditions, using 253 fs pump pulses in both cases, with two detection-crystal thicknesses: the 3-mm-thick GaP crystal used for the spatially integrated measurements and a 0.5-mm-thick GaP crystal. The dotted curves in Fig. **2**b show the calculated EOS response functions for 0.5-, 3- and 5-mm-thick GaP crystals. The 5-mm response is included for comparison with the wide-field EO imaging configuration described below; the conventional EOS spectra in this panel were measured only with the 0.5- and 3-mm crystals. Detection is modelled by the EOS response function of the GaP crystal[29], which includes the mismatch between the THz phase velocity and the group velocity of the probe pulse, the THz absorption of GaP, the frequency dependence of the electro-optic coefficient and the finite probe pulse duration; for the thicker crystals, the dispersion of the probe pulse inside the crystal is also taken into account. The calculated responses predict stronger high-frequency attenuation for the thicker crystals, whereas the measured 0.5- and 3-mm spectra have similar spectral shapes and reach the noise floor at approximately the same frequency. This agreement indicates that thickness-dependent filtering by the 3-mm GaP crystal is not the dominant cause of the observed roll-off near 3 THz. The comparison does not exclude limits shared by both measurements, including the probe-pulse duration, collection geometry and finite DR. The 3-mm crystal was therefore retained for the conventional EOS measurements because of its higher signal-to-noise ratio. The two measurements are shown individually, in the time and the frequency domain and together with the calculated crystal responses, in Supplementary Note 2 and Fig. S1.

The low-frequency-weighted spectrum of the emitter reflects both the relatively long carrier lifetime in SI-GaAs and, importantly, the large-area geometry of the source. Because the emitting aperture is 10 × 10 mm² and the collection optics are designed for this large area, the strongly divergent low-frequency components, which are commonly lost in emitters with small excitation spots or substrate lenses, are collected efficiently here. In this sense the spectral weight at low frequencies is largely a consequence of efficient collection rather than of a loss of high-frequency content. The detected pulse shape and spectral weighting therefore also depend on the collection and detection geometry. Further factors are expected to shift weight away from the high-frequency part of the spectrum: the 10 μm electrode separation, the use of SI-GaAs rather than LT-GaAs, and excitation high in the conduction band at 515 nm. These contributions cannot be separated by the present measurements alone.

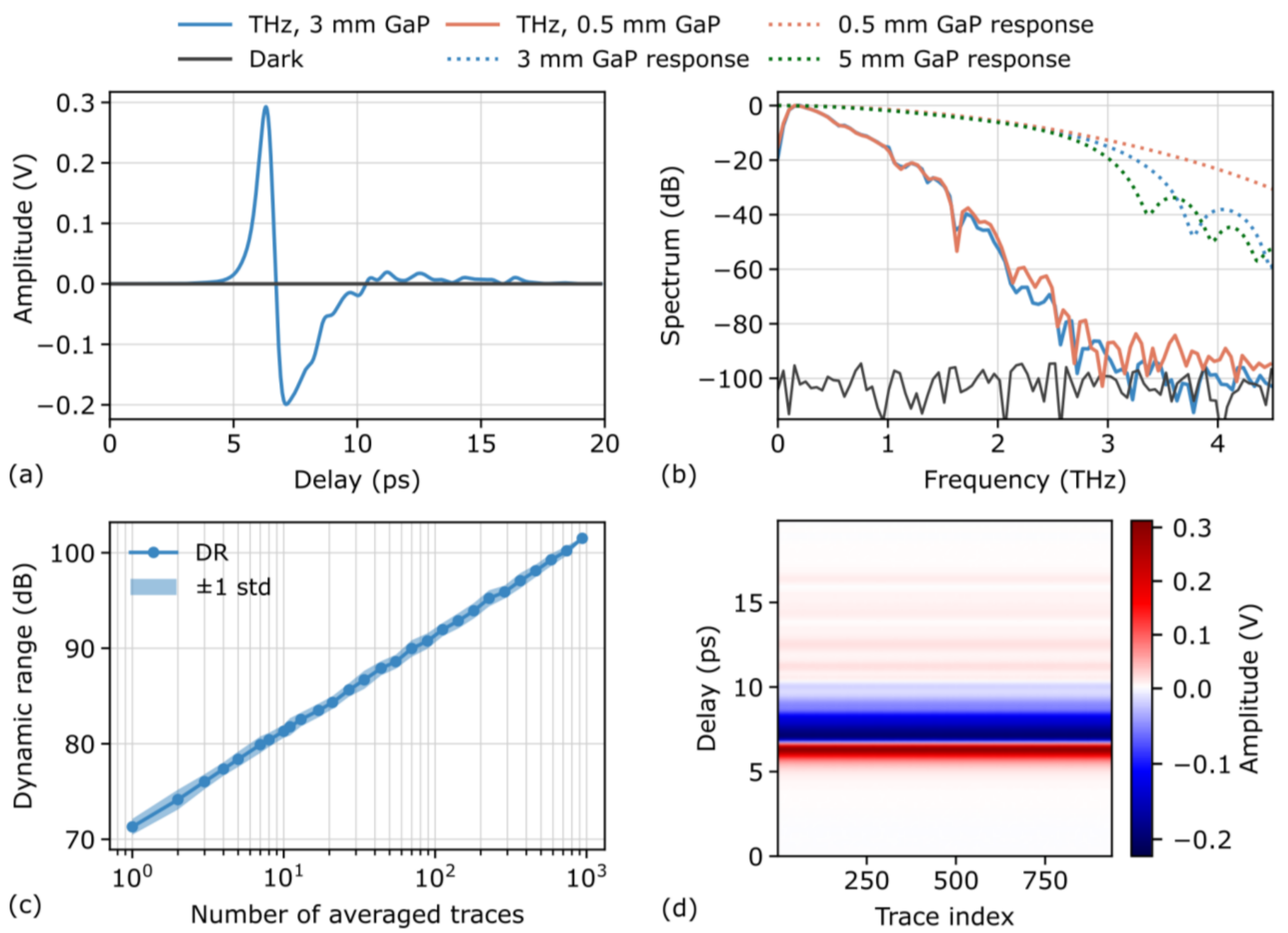

**Fig. 2** | THz waveform**, spectrum, and dynamic range characterization. a** THz electric field measured by electro-optic sampling (EOS) in a 3-mm-thick GaP crystal, at 12.6 W of optical pump power and 84 kV/cm acceleration field, with the corresponding dark trace recorded with the bias voltage switched off; at this bias the emitter delivers 34.6 ± 0.2 mW of THz average power, below the 38.3 mW record obtained at 95 kV/cm. **b** THz spectra recorded with 3-mm-thick and 0.5-mm-thick GaP detection crystals under identical excitation with approximately 253 fs green pump pulses, each normalized to its peak. Dotted curves show the calculated EOS responses of 0.5-mm (orange), 3-mm (blue) and 5-mm (green) GaP crystals. The 5-mm response is shown for comparison and corresponds to the crystal used in the wide-field THz imaging measurements of Fig. 3; no measured 5-mm spectrum is included in this panel. The dark trace corresponds to the 3-mm crystal. **c** Peak dynamic range (DR) evolution estimated by random-subset averaging; symbols show the mean and the shaded region the standard deviation. **d** Individual THz traces versus trace index, showing stable timing and amplitude during acquisition.

## 2. Wide-field electro-optic imaging setup

To exploit the high average power and high repetition rate of the source, we combine it with parallel EO imaging for rapid, spatially resolved THz-TDS. The experimental setup is comparable to that of Fig. **1**a, with a modified EOS detection arm as shown in Fig. **3**a. In this implementation, a quarter-wave plate (QWP) and a Wollaston prism (WP) convert the polarization modulation into an intensity difference between two spatially separated probe images with orthogonal polarizations that are imaged simultaneously on a fast camera. To preserve the spatial correspondence between the EO crystal plane and the sensor of the camera, a 4f-imaging system with unit magnification, consisting of two identical lenses of focal length $f = 150$ mm, relayed the crystal plane onto the camera detector. We selected a 5-mm-thick GaP crystal because it had the best optical quality over the required aperture among the crystals available in our laboratory and provided a strong EO signal. Optical uniformity is particularly important for wide-field detection, where local damage, scattering centers and surface imperfections can appear as image artifacts. This choice trades spatial resolution and high-frequency response for sensitivity over the field of view. Its calculated EOS response is included in Fig. **2**b, and the resulting bandwidth is discussed below. The estimated propagation-limited THz spatial resolution ranges from ~1.3 mm at 0.1 THz down to ~300 µm at 2 THz. We discuss this aspect in more detail in the Discussion section.

To improve the spatial uniformity of the probe beam over the THz spot in the GaP crystal, the probe was expanded and diverging at the crystal, with its diameter adjusted to approximately 4 mm ($1/e^2$ intensity), approximately twice the THz beam diameter at the EO crystal. To compensate for the resulting reduction in probe intensity, the fraction of laser power directed to the probe arm increased from 1% to 5% at the beam splitter.

For the imaging experiment, a global shutter high-speed camera (FASTCAM Mini WX100, Photron) operated at 6250 frames per second was used. During acquisition, the probe beam remained continuously incident on the camera, while the THz waveform was sampled by a continuously oscillating delay line operating at 1 Hz over a 20-ps scan range. Each oscillation contains a forward and a backward sweep, giving a waveform acquisition time of 500 ms per sweep. The 1 Hz scan rate was chosen to match the camera buffer capacity. Each continuous acquisition run is limited to approximately 7 s (12–13 waveforms), after which the buffer must be read out (ping-pong buffering for maximized throughput). A 7 s acquisition, therefore, contains approximately 12–13 THz waveforms and corresponds to more than 43,000 recorded frames (512 x 1024 pixels, 16-bit depth). To map each recorded frame to the corresponding THz–probe delay, a separate digitizer (Spectrum Instrumentation, Model DN2.442-04), gated during active recording to the internal camera buffer, acquired the analog shaker position signal alongside the exposu re trigger for each frame the camera acquired. In post-processing, the shaker output was used to determine the delay value associated with each frame, allowing reconstruction of the THz waveform along the temporal axis. The images were then separated into the two WP polarization channels, cropped into the relevant regions of interest (ROI), spatially registered, and analyzed pixel by pixel. The details of post-processing of the data are discussed in the Methods section.

To benchmark the camera-based EO imaging against conventional detection, the same THz pulse was characterized in three configurations: a single 10 µm camera pixel at the beam center (Fig. **6**a–d; additional positions in Fig. **3**c are characterized in Supplementary Note 6 and Fig. S6), the spatial average over the full probe aperture (**Fig. 6**e–h), and conventional balanced photodetection EOS obtained by replacing the camera with a balanced photodetector and lock-in detection, with the same expanded probe and 5-mm GaP crystal (**Fig. 6**i–l). A single trace corresponds to one 500 ms delay sweep. For the camera measurements, 100 waveforms were accumulated from eight buffer-limited runs of approximately 7 s each, giving approximately 56 s of active recording.

As shown in Fig. 3b, the two spatially separated probe images produced by the WP are recorded simultaneously on the sensor; the rectangular ROIs define the left (L) and right (R) detection channels used for channel separation, image registration, and balanced-signal evaluation. The registered, common field of view of the two channels is shown in **Fig. 3**c, together with the representative pixel positions used for the spatially resolved analysis below; the white outline marks the intensity mask (pixels above 10% of the peak intensity) used to exclude low-intensity pixels from the α-map calculation. In Fig. 3d,e, we compare EOS traces obtained with the camera and with a balanced photodetector using the same in imaging geometry. The whole-area camera-averaged spectrum in Fig. 3e has a lower upper-frequency limit (~2 THz) than the conventional EOS spectrum in Fig. 2 (~3 THz), consistent with the different detection geometry and DR used for wide-field imaging (see Discussion). A single 10-µm pixel recovers a clean single-cycle THz waveform in one 500 ms trace, with a peak DR of approximately 26 dB. Averaging raises the DR to approximately 50 dB over 100 traces. The whole-area camera measurement reaches approximately 73 dB, compared with approximately 90 dB for balanced-photodetector EOS under the same imaging geometry. The waveforms and spectral features agree over their common measurable frequency range, while their usable bandwidths depend on the respective noise floors (Methods).

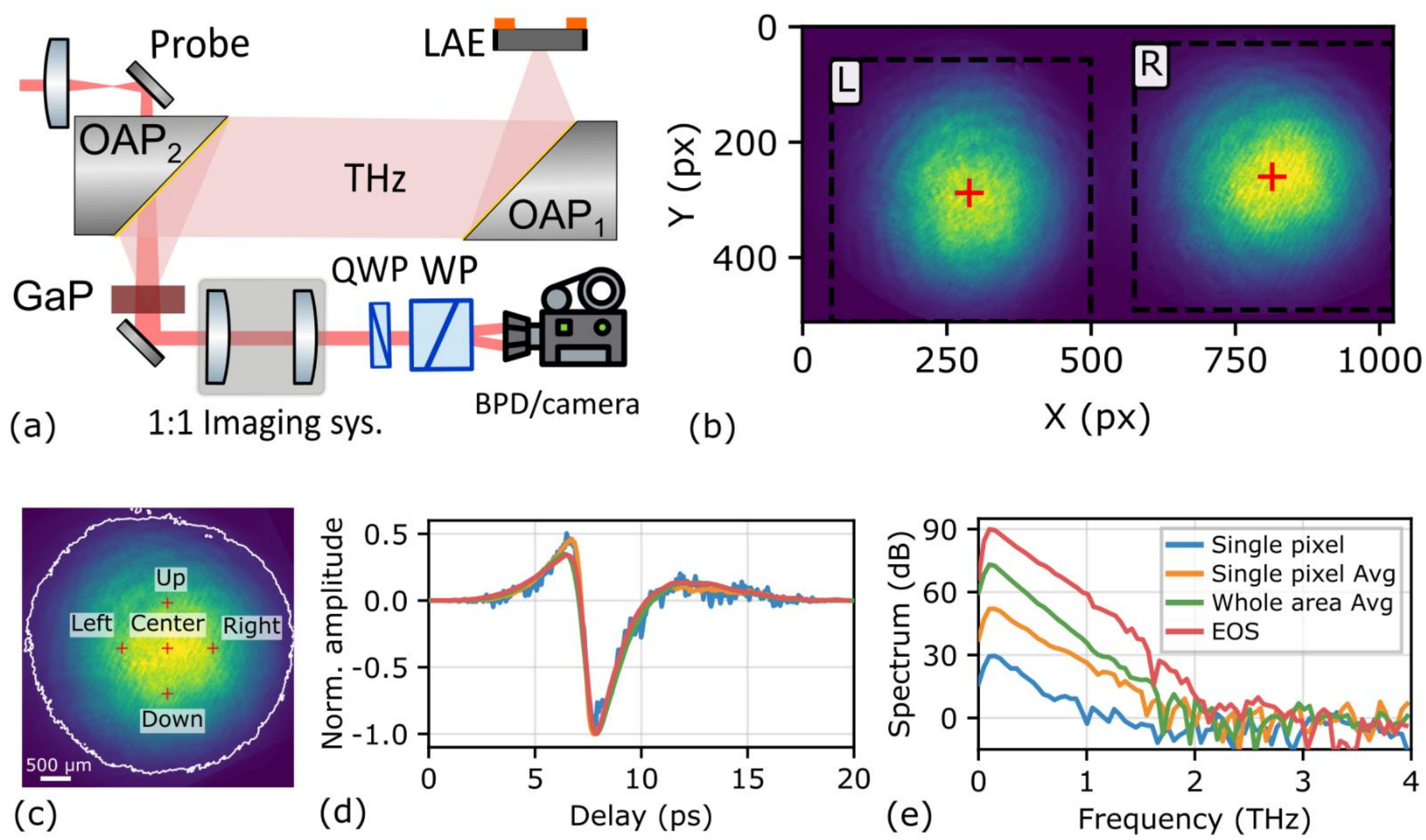

**Fig. 3 | Parallel electro-optic imaging detection. a** Schematic of the modified THz detection arm. **b** Single camera frame with the left (L) and right (R) regions of interest (ROIs) marked. **c** Registered probe field of view with the representative pixel positions; the white outline marks the intensity mask (pixels above 10% of the peak intensity) used to exclude low-intensity pixels from the α-map calculation. Scale bar: 500 µm. **d** Normalized THz waveforms: a single 10 µm camera pixel (single trace), its 100-trace average, and the spatial average over the whole aperture, compared with balanced-photodetector EOS measured using the same expanded probe and 5-mm GaP crystal. **e** Corresponding power spectra for the single pixel, single-pixel average, whole-area average, and EOS detection.

#### Spatiotemporal reconstruction of the THz field

To illustrate the spatiotemporal capability of the parallel detection scheme, **Fig. 4** reconstructs the complete THz field $E(x, y, \tau)$ from the imaging dataset. Because every camera pixel records a full temporal waveform simultaneously, the data recorded forms a three-dimensional field map that can be viewed either as a sequence of two-dimensional spatial snapshots at fixed delay or as spatiotemporal sections through the beam. In this way, we evaluate the overall performance of the reconstructed electric field, which allows us to understand the limitations of the system in view of future applications in imaging. **Fig. 4**a shows snapshots of the transverse field at selected delays spanning the single-cycle THz transient (a corresponding video of the field evolution is provided in the supplementary material). The transverse field distribution therefore does not simply scale up and down with delay; its spatial shape evolves through the cycle, which is captured simultaneously across the field of view during the delay sweep.

To examine the spatiotemporal structure more directly, **Fig. 4**b,c show the field along orthogonal cuts through the beam center, $E(x,\tau)$ at fixed $y_0$ and $E(y,\tau)$ at fixed $x_0$, plotted as three-dimensional surfaces, with selected individual pixel waveforms extracted in **Fig. 4**d,f for clarity. The waveforms retain a similar single-cycle shape across the field of view, while their amplitudes decrease toward the beam edges. Their arrival times also vary with position, as discussed below; the measurements therefore support preservation of the local waveform shape without implying the absence of spatiotemporal structure. In the two-dimensional projections (**Fig. 4**e,g), the sharp zero-crossing between the positive and negative half-cycles provides a tracer of the local THz–probe relative arrival time. The half-cycle bands are not perfectly horizontal: the field arrives at a slightly later delay near the beam center than at the edges, by approximately 0.6 ps across the aperture, with the same sense and comparable magnitude along both $x$ and $y$ cuts. This symmetric delay is consistent with a weak curvature of the THz–probe relative arrival-time surface at the detection plane. Because the probe is diverging, the curvature cannot be assigned to the THz field alone without separately characterizing the probe. The same structure is visible in **Fig. 4**b,c.

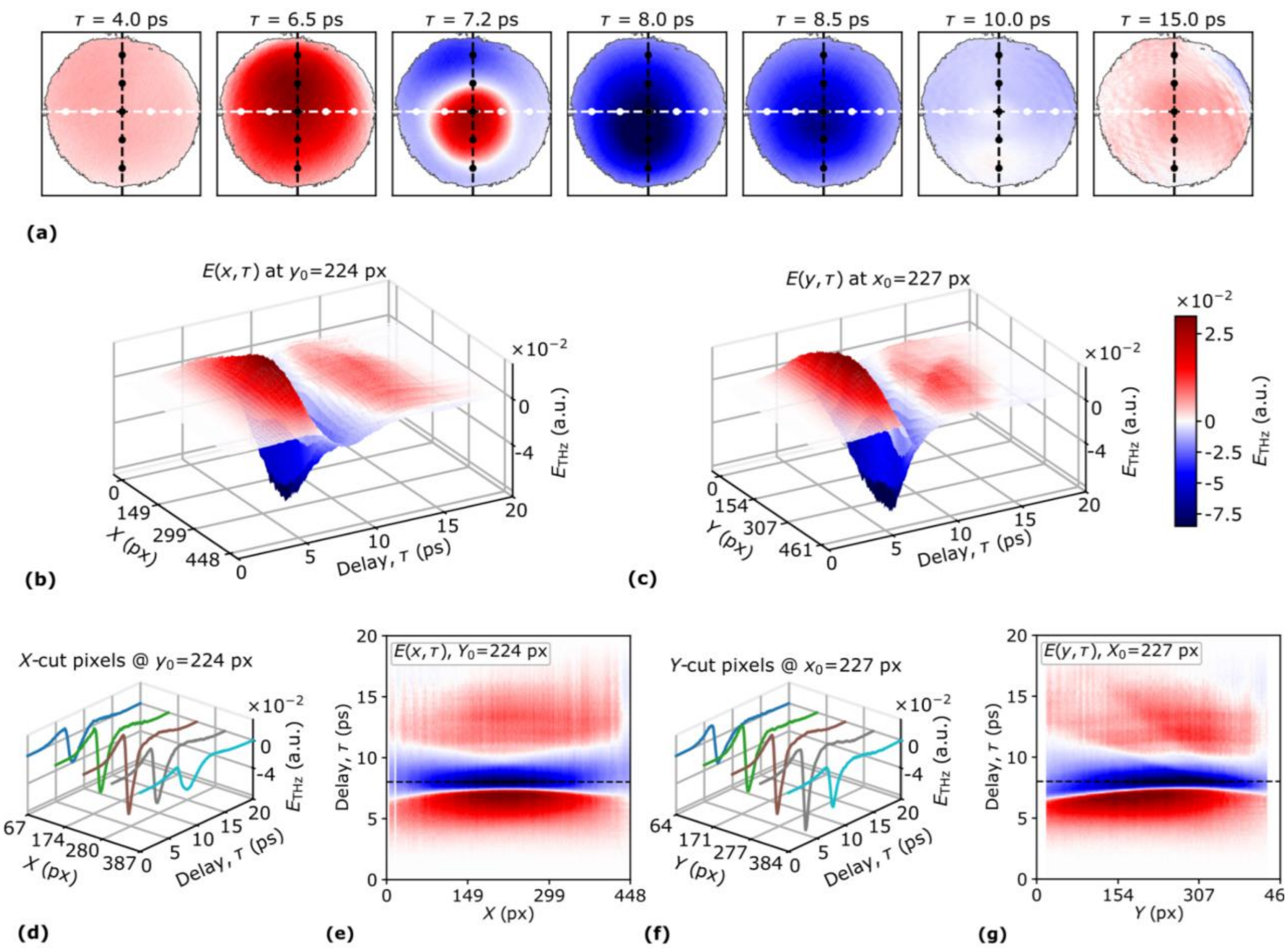

**Fig. 4 | Reconstructed spatiotemporal THz field from a parallel-imaging dataset. a** Transverse field snapshots at selected delays across the single-cycle transient; dashed lines mark the x and y cuts used below. **b**, **c** Field along orthogonal cuts through the beam center, $E(x,\tau)$ at fixed $y_0$ and $E(y,\tau)$ at fixed $x_0$, plotted as 3D surfaces. **d**, **f** Selected individual pixel waveforms along the x and y cuts. **e**, **g** Two-dimensional projections of **b**, **c**, with the field amplitude encoded in color; the dashed line marks the delay of peak field, $\tau \approx 8$ ps, defined as the maximum of $|E(x,\tau)|$ over the full frame. Field values are in arbitrary units throughout.

## Discussion

### High power large area emitter

In our experiment the emitter reaches a conversion efficiency of 0.3%, whereas standard PCAs reach several percent in optimized cases. One reason is that only 20% of the incident optical power is absorbed in the active photoconductive regions of the LAE, since the metallization structure covers ~75% of the emitter surface. Fresnel losses at the air–GaAs interface reduce the efficiency further[26,28]. The first and more important limitation can be compensated by selectively focusing the pump onto the exposed photoconductive regions, for example using a microlens-array approach[30,31]. Rescaling the measured efficiency by the estimated 20% active absorption gives approximately 1.5%, provided the conversion per absorbed optical power is maintained; this is an estimate rather than a prediction that includes saturation. Higher efficiencies up to 3.4% have been reported in other photoconductive devices[32], however these are not in comparable geometries or materials to our result.

The spectrum measured by conventional EOS with a probe focused to ~400 µm in the 3-mm-thick GaP crystal (Fig. 2b) extends to ~3 THz. The wide-field imaging measurements of Fig. 3 used an expanded, diverging probe with a diameter about twice that of the THz beam to illuminate the field of view approximately uniformly, together with a 5-mm-thick GaP crystal. The balanced-photodetector reference in **Fig. 3**d,e was measured with the same probe and crystal geometry as the camera measurements. This comparison therefore tests the two readout methods under a common detection geometry.

These changes in detection geometry can alter the measured spectral weighting. For comparable incident THz beam profiles, higher-frequency components form smaller focal spots and have shorter confocal ranges. The focused probe in Fig. 2 samples the central region of the THz focus, whereas the expanded probe and whole-area averaging in Fig. 3 can give greater relative weight to spatially extended, lower-frequency components. Integration through the thicker crystal can further modify this weighting through longitudinal overlap and the frequency-dependent EO response (the calculated 5-mm response is shown in Fig. 2b). Spatial averaging can also suppress high-frequency components when local waveforms have different arrival times, as observed in Fig. 4. The response curve alone does not include these spatial-averaging effects. These effects, together with the lower DR of the whole-area camera signal (Methods), provide a plausible explanation for its upper-frequency limit of ~2 THz, compared with ~3 THz in Fig. 2. The ~1.5 THz limit of a single-pixel measurement is also influenced by its lower signal-to-noise ratio. The observed cutoffs therefore characterize the respective detection configurations and do not directly establish the intrinsic bandwidth of the emitter.

This low-frequency spectral range can be useful where water absorption is a concern, whereas other applications require broader bandwidths. Optimized photoconductive materials with faster carrier recovery times, combined with bandgap engineering to better match 1030 nm, could deliver both broader bandwidths and higher efficiency, and ongoing work on alternative material platforms is promising in this respect[33–35]. We also note that for the imaging application targeted here, other high-power THz generation mechanisms — organic crystals[22], or plasma-based sources[36]— offer a potential path to higher bandwidths provided the electro-optic imaging setup also supports the bandwidth.

### Electro-optic imaging

**Resolution** Our data demonstrates that the combination of high-power sources and parallel EO imaging enable rapid, spatially resolved THz imaging, and comparison with conventional single-pixel EOS confirms that the parallel approach reproduces the same waveform shape and bandwidth across the full field of view. When assessing the applications of such a fast spectroscopic THz camera, the achievable spatial resolution is a key consideration. For sample-based near-field imaging in this configuration, diffraction of the THz field within the 5-mm-thick GaP crystal is expected to dominate over the camera sampling pitch and the optical imaging resolution: the field distribution imprinted by the sample spreads as it propagates through the crystal before being sampled by the probe[37]. Following the diffraction criterion of Doi et al.[37], the propagation-limited resolution is $\delta x \approx 0.61\,(\lambda L/n)^{1/2}$, with $\lambda$ the free-space THz wavelength, L the crystal thickness and n the THz refractive index of GaP. For L = 5 mm this gives an estimated resolution of ≈1.3 mm at 0.1 THz, improving to ≈0.4 mm at 1 THz and ≈0.3 mm at 2 THz.

Reaching a pixel-limited resolution of 10 µm would require thinning the crystal to a few micrometres at the upper end of our spectrum (≈3 µm at 1 THz, ≈6 µm at 2 THz), and to sub-micrometre thickness at 0.1 THz. Near-field EO imaging with micrometre-thick $LiNbO_3$ crystals has demonstrated 5–9 µm (λ/600) resolution at 100 GHz [16], confirming that this regime is experimentally accessible. Such a thousandfold reduction in thickness would, however, significantly reduce the dynamic range at constant field of view. Some applications may tolerate this, however for cases where the loss cannot be tolerated, this can be partially recovered either by increasing the power of the source if available or by concentrating the available power onto a smaller area at the expense of the wide-field. A new compromise between crystal thickness, which sets both resolution and detection bandwidth, field of view and dynamic range must therefore be found for a given application. High THz average power becomes the main criterion, allowing thinner crystals or larger areas at a dynamic range acceptable for a given application.

In future systems where higher spatial resolution may be desired, several additional measures can recover part of the DR sacrificed by thinner crystals. Reflection-mode probing with a high-reflection-coated crystal surface doubles the effective interaction length at fixed thickness[37]. Spectral filtering of the optical probe before and after the EO crystal has been shown to triple the EO modulation depth while simultaneously extending the detected bandwidth [17]. Finally, materials with a higher electro-optic coefficient than GaP — thin $LiNbO_3$ ($r_{33} \approx 31$ pm/V) as used in near-field EO microscopy [17], or organic EO crystals with even larger coefficients [22], can recover an order of magnitude or more in signal, provided phase-matching and absorption at the THz wavelength are managed. Together with the extensive waveform averaging enabled by high repetition rates and fast delay scanning, these measures define the parameter space in which a next-generation system can be optimized.

**Faster detection** The acquisition rate of full time-domain image stacks is set by the oscillating delay line. The 1 Hz scan frequency balances the available camera frame rate, buffer capacity and required THz bandwidth. Increasing the shaking frequency is in principle possible, multiplying the trace rate directly, but the camera must follow: for a sinusoidal delay covering a time window $T_w$, the sweep velocity peaks at the center of the trace, and alias-free sampling up to $\nu_{max}$ requires $f_{cam} \geq 2\pi \cdot f_{shake} \cdot \nu_{max} \cdot T_w$ [25,38], resulting in compromises to be considered regarding bandwidth and available camera frame rates.

**Data acquisition and acceleration** Beyond speed and DR, a practical consequence of recording the full THz waveform in parallel at every camera pixel is the large data volume generated. Acquiring all frames at the full frame rate and bit depth, including the true noise floor via two dark measurements[39], produced raw data on the order of 700 GB for a single experiment (56 s of averaging), creating bottlenecks in data transfer (here set by the camera-to-PC Ethernet link), storage, and pixel-by-pixel post-processing. In future implementations, this data-handling overhead, rather than the optical or THz hardware, is likely to set the practical limit for scaling the field of view, pixel count, or waveform length.

Several technical routes can address this overhead and accelerate the platform further. First, direct streaming to a binary format such as HDF5 could combine region-of-interest cropping, calibrated registration, balanced subtraction and real-time waveform accumulation. Storing accumulated waveforms would reduce the number of individual frames written to disk. Second, the inherent sparsity of the THz signal can be exploited: sparse or adaptive delay sampling reduces the number of temporal samples when only a restricted spectral range or a known waveform structure is needed, as recently demonstrated by compressed-sensing reconstruction of field-resolved traces beyond the Nyquist frequency[40], while spatial compressive schemes reduce the number of points sampled across the field of view[12], together offering substantial reductions in both acquired data and processing load. Finally, synchronous detection schemes such as dynamic-subtraction or lock-in-type camera readout, synchronized to a modulation of the THz beam, are known to suppress technical noise in camera-based EO imaging[25,41] and would narrow the remaining gap between the whole-area camera signal and conventional lock-in EOS, while also reducing the averaging, and hence the data, required to reach a given dynamic range. Together with higher THz average power, these improvements provide a clear path toward thinner EO crystals, larger fields of view, faster waveform acquisition and broader-bandwidth parallel THz imaging.

Broader-bandwidth operation requires adapting both the source and the EO detector. Crystal thickness must account for THz–probe velocity mismatch and absorption. At our 5 mm thickness, phase mismatch and THz absorption in GaP progressively attenuate the upper spectral components, so the thinner crystals motivated above by spatial resolution would simultaneously flatten and extend the detection response. Furthermore, temporal sampling is another critical consideration: the required frame rate increases linearly with the target bandwidth, so that capturing, for example, 5 THz over a 10 ps window at 20 Hz shaking requires ~6.3 kfps, rising to ~12.6 kfps at 10 THz over the same window — still within reach of the present camera class.

In conclusion, we have demonstrated fast, wide-field terahertz imaging in which the full THz waveform — amplitude and phase — is recorded in parallel at every pixel, enabled by a record-high average power photoconductive large-area emitter. Driven by a commercial, industrially robust Yb-based femtosecond laser, the source delivers a record-high 38.3 ± 0.2 mW of THz average power at a 400 kHz repetition rate, sufficient to feed a parallel electro-optic detection scheme across a wide field of view. With this approach we reconstruct complete single-cycle THz waveforms across ~145,000 pixels spanning a ~14.5 mm² field of view in 500 ms, corresponding to ~$2.9 \times 10^5$ spatially and spectrally resolved pixels per second at the 10-µm camera sampling pitch during active acquisition. This rate counts sampled pixel waveforms and is not normalized to the independent spatial resolution. These results establish that high THz average power is the enabling resource for practical parallel electro-optic imaging, eliminating the need for specialized kHz-laser. The routes to higher resolution, faster acquisition and broader bandwidth discussed above could extend high-average-power THz-TDS as a practical platform for high-speed, material-resolved inspection and for the spectroscopic study of irreversible, second-timescale processes such as hydration changes in biologically relevant systems.

## Methods

### Power measurements

The THz average powers reported in the Results were measured by replacing the EO detection crystal with a calibrated pyroelectric THz power meter (SLT GmbH, *THz20*, calibrated by the German Metrology Institute PTB). To suppress pump leakage, two sheets of 180-µm-thick white paper were inserted as THz-pass filters in the collimated THz beam path and in front of the power meter, while the transparent protective cover attached to the sensor head shields the detector and reduces the influence of ambient air fluctuations. Their transmission (88% THz power transmission for each paper and 92% for transparent protective cover) was measured independently and considered in the reported power values, with a combined transmission factor of $0.92 \times 0.88^2 \approx 0.71$. When either the optical pump or the electrical bias was switched off, the measured signal dropped to the noise floor, confirming that the recorded THz power was free of detectable pump leakage and spurious electrical heating contributions from the emitter. The calibrated THz power meter requires a modulated THz signal at 18 Hz with 50% duty cycle. Since we modulate the pump beam and this may affect the output, we compared two schemes: chopping the optical pump beam and chopping the THz beam after generation. Both methods yielded the same THz power, indicating that the increased input average power without optical chopping does not significantly degrade efficiency under these conditions. For convenience, the optical pump beam was chopped in all subsequent power measurements.

**Repetition rate dependent operation of the LAE**

As part of the emitter characterization, we investigated the influence of the laser repetition rate by varying it from 50 kHz to 400 kHz while keeping the pump spot diameter fixed at 7 mm and the acceleration field at 55 kV/cm. In each case, the PWM bias was synchronized to the corresponding laser repetition rate. Because the repetition rate sets the pulse energy at a given average power, this sweep covers more than one order of magnitude of pump pulse energy, from 1.25 µJ at 400 kHz to 30 µJ at 50 kHz, so that the meaningful comparison is the THz pulse energy as a function of the pump pulse energy rather than of the average power. Plotted in this way, Fig. 5c shows that the seven repetition rates, each covering a different window of pulse energy, together trace one common curve, which rises nearly quadratically at low pulse energy and approaches a linear dependence at the higher pulse energies.

This data are analyzed in detail in Supplementary Note 5. As shown there, the low-energy behavior is the quadratic scaling expected in the unsaturated limit, and the local exponent $d \ln E_{THz} / d \ln E_{opt}$ falls from 1.9 to below one as the pulse energy increases, following the same path for every repetition rate. The roll-off is therefore predominantly governed by excitation fluence and is consistent with the carrier-screening mechanisms discussed below. The repetition rate itself leaves only a small residual effect: at equal pulse energy, a fourfold higher repetition rate costs 13–16% in THz pulse energy, consistent with the larger average power, and therefore the higher heat load, carried by the device. The dominant fluence dependence suggests that lower-repetition-rate operation could increase the THz pulse energy and peak field, subject to the residual repetition-rate dependence and the accessible pump-pulse energy.

**Power and Bias Optimization**

To optimize the excitation conditions, we first studied the THz yield as a function of optical pump power, pump spot size and acceleration field. The optimum performance was obtained for a pump spot diameter of 7 mm ($1/e^2$ intensity), corresponding to 99% of the optical power incident on the active area, at a maximum pump power of 12.6 W. Fig. **5**a shows the measured THz average power as a function of pump power at three bias voltages of 20, 51 and 75 V, corresponding to acceleration fields of 20, 51 and 75 kV/cm, under synchronized unipolar PWM bias operation at 400 kHz. To supply the photocurrent required in this high-power excitation regime, and in the subsequent measurements, two identical voltage amplifiers (Falco Systems WMA-300) were connected in parallel to the waveform generator, as the current output of a single amplifier is limited to 300 mA. The inset in Fig. 5b shows the THz beam profile recorded by a THz camera (Swiss Terahertz, *Rigi S2*) with a focused beam diameter ($1/e^2$ intensity) of 1.86 mm × 1.83 mm from a two-dimensional Gaussian fit, exhibiting a close-to-ideal Gaussian shape.

At all three acceleration fields the THz average power initially grows almost quadratically with the optical pump power, as expected in the unsaturated limit where the radiated THz field is proportional to the photo-induced current and hence to the photogenerated carrier density. Towards higher pump powers the scaling becomes sub-quadratic, consistent with screening of the applied bias field by photogenerated carriers. Two contributions are well established for this device class: the space-charge field that builds up as electrons and holes separate within the 10 µm gaps, and the near-field (radiation) screening by the emitted THz field itself, which is characteristic of large-aperture emitters[42–44].

The onset of this roll-off shifts systematically to higher pump power as the acceleration field is increased, from approximately 0.9 W at 20 kV/cm to 1.8 W at 51 kV/cm and 2.4 W at 75 kV/cm, corresponding to peak pump fluences of approximately 12, 23 and 31 µJ/cm². Both mechanisms scale with the photogenerated carrier density and are therefore governed by the pump fluence; the shift of the onset with the applied field is analysed in Supplementary Note 7. The same trend has been reported for this emitter type at lower acceleration fields and lower repetition rate[28]. As a direct consequence, the pump power at which the optical-to-THz conversion efficiency peaks moves from 4 W at 20 kV/cm to 10 W at 51 kV/cm, while at 75 kV/cm the efficiency remains at its maximum up to the highest available pump power of 12.6 W. Operating at a high acceleration field therefore both increases the THz output directly and keeps the emitter in its quasi-linear regime over the full available pump power. The three data sets are analysed individually in Supplementary Note 4, where the local scaling exponent, the definition of the roll-off onset and the associated uncertainties are given in detail.

Next, we investigated the influence of the acceleration field on the LAE performance while keeping the pump power fixed at its maximum value of 12.6 W, as shown in Fig. **5**b. Synchronized pulsed-bias operation allowed us to apply acceleration fields of up to 95 kV/cm, which is the main driver of THz generation efficiency in these photoconductive devices. This was achieved by using a waveform generator phase-locked to the laser, together with two parallel voltage amplifiers (Falco Systems WMA-300), to apply a unipolar PWM bias with 650 ns voltage pulses (duty cycle 26%), synchronized to the 400 kHz laser repetition rate. Under these conditions, a record-high THz average power of 38.3 ± 0.2 mW was obtained, corresponding to an optical-to-THz conversion efficiency of 0.30%. High-power operation was verified on different LAE devices, for which comparable maximum THz powers were obtained under the same excitation conditions, indicating that the result is reproducible and not specific to a single sample.

The shape of the bias dependence in Fig. 5b is in itself informative. Below approximately 40 kV/cm the emitted THz field grows almost linearly with the applied acceleration field (i.e. the average power quadratically), as expected for bias-driven acceleration of the photocarriers. Above this value the increase becomes progressively sub-linear, and the curve flattens above ~90 kV/cm: raising the acceleration field by a further 4%, from 91 to 95 kV/cm, yields only 2% more THz power. This roll-off is consistent with the transfer of electrons from the central Γ valley into the L and X satellite valleys of GaAs, where their much lower mobility reduces the transient photocurrent[28]; excitation at 515 nm, which already places carriers well above the conduction-band minimum, can facilitate intervalley transfer. Together with the fluence dependence, these measurements show that the highest available bias yields the largest THz power while maintaining efficient conversion at full pump power. A quantitative comparison of both dependences with photocurrent transport models is given in Supplementary Note 7. The way the emitted waveform and its spectrum evolve when the pump power and the acceleration field are varied independently over the full accessible range is examined in Supplementary Note 3. Both parameters drive the emission towards the same limiting waveform: at low excitation the transient is longer and the spectrum peaks near 0.12 THz, whereas raising either the pump power or the bias shifts the spectral peak to about 0.20 THz and roughly doubles the half-maximum frequency, after which the normalized waveform becomes insensitive to a further increase of either parameter, and the amplitude is the principal remaining change.

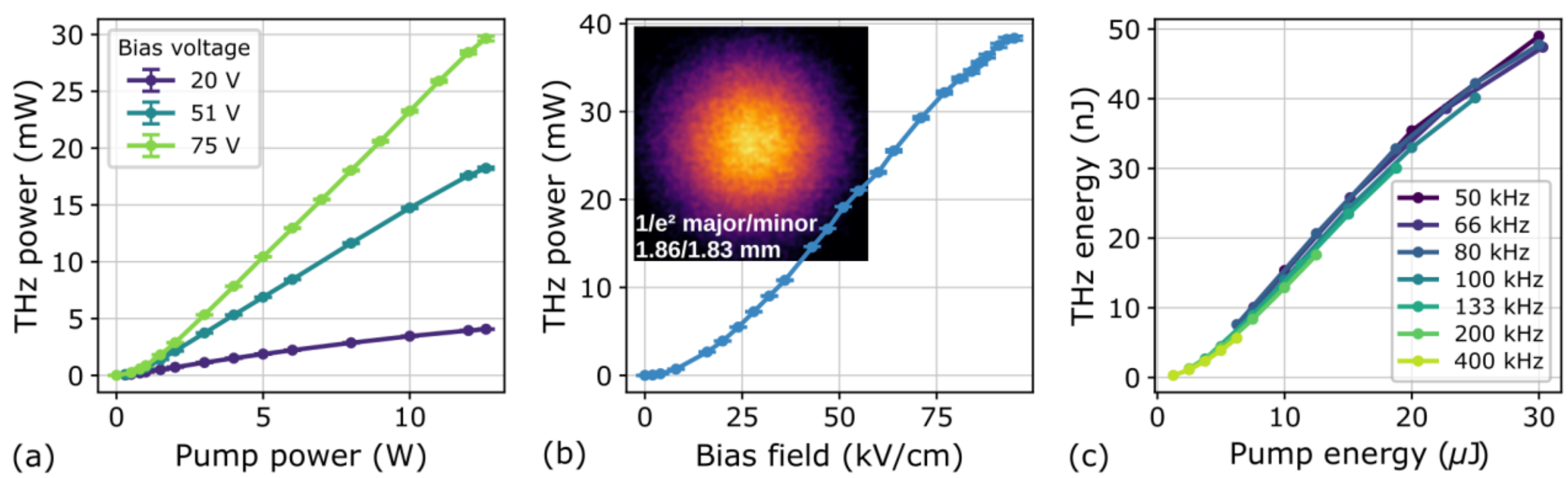

**Fig. 5 | THz power and emitter optimization.** All measurements were taken with a 7 mm ($1/e^2$) pump spot on the emitter. **a** THz average power versus optical pump power at 400 kHz for three bias voltages of 20, 51 and 75 V, corresponding to acceleration fields of 20, 51 and 75 kV/cm; the onset of the roll-off shifts to higher pump power as the acceleration field increases. **b** THz average power versus acceleration field at a fixed optical pump power of 12.6 W and 400 kHz, under synchronized unipolar pulsed-bias operation, reaching 38.3 ± 0.2 mW at 95 kV/cm. Inset: THz beam profile at focus. **c** THz pulse energy versus optical pump pulse energy for different laser repetition rates, at an acceleration field of 55 kV/cm.

### Data acquisition, image registration, and signal processing

Because the two polarization images must represent the same spatial position in the EO crystal, image registration was calibrated using a rigid transformation determined from a 1951 USAF resolution target placed at the crystal plane. The same target was used to validate the effective magnification of the imaging system. After registration, the raw balanced EO signal for each pixel was obtained from the difference between the left and right polarization channels. To compensate for residual imbalance between the two detection arms, a balancing factor α was determined from reference measurements with the THz beam blocked: either as a scalar from the ratio of integrated probe intensities in the two channels, or as a spatially dependent map $\alpha(x,y)$ from the ratio of the registered, averaged reference images. The corrected EO signal was then evaluated as $I_L - \alpha I_R$ (scalar) or $I_L - \alpha(x,y) I_R$ (pixel-wise).

Since the probe beam has a Gaussian transverse intensity profile, the raw differential EO signal still contained a residual spatial weighting by the local probe intensity[18]. This was removed by normalizing the corrected differential signal to the corresponding reference intensity. Specifically, the normalized signal was calculated as $(I_L - \alpha I_R)/(I_{L,\mathrm{ref}} + \alpha I_{R,\mathrm{ref}})$ for scalar balancing, or $(I_L - \alpha(x,y) I_R)/(I_{L,\mathrm{ref}} + \alpha(x,y) I_{R,\mathrm{ref}})$ for pixel-wise balancing. This procedure cancels the spatial probe envelope, so that the recovered signal represents the THz-induced EO modulation. Normalization is particularly important for quantitative comparison between different pixels and ROIs because it suppresses artificial amplitude variations caused by probe-beam inhomogeneity and improves the robustness of the reconstructed THz traces [18].

### Validation of parallel electro-optic imaging against conventional balanced-detection

A single 10-µm pixel recovers a clean single-cycle THz waveform in one 500 ms trace, with a peak DR of approximately 26 dB (**Fig. 6** a–c). Averaging raises the DR to approximately 50 dB over 100 traces. Importantly, most of the DR improvement is concentrated in the first approximately 20 traces. The trace-index maps (**Fig. 6**d,h,l) show no significant timing drift or amplitude degradation across the acquisition, confirming that the DR accumulates from reproducible waveforms rather than occasional favorable traces.

Spatially averaging over the full probe aperture increases the DR to approximately 73 dB (**Fig. 6**e–g), while balanced-photodetector EOS with the same expanded probe and 5-mm GaP crystal reaches approximately 90 dB under otherwise identical conditions (**Fig. 6**k). This ordering reflects the detected probe-photon and THz-field budget per detection element: a single camera pixel integrates only a small fraction of the available probe light, the whole-area sum recovers essentially the full photon budget, and the photodetector additionally benefits from synchronous lock-in detection that simple frame-averaging does not employ.

The camera and photodetector measurements have similar waveform shapes and spectral features over their common measurable frequency range (**Fig. 6**a,b,e,f,i,j). In the time domain (**Fig. 6**a,e,i), each waveform is independently normalized to its own peak amplitude; in the frequency domain (**Fig. 6**b,f,j), each spectrum is normalized to its own noise floor, so that the spectral shape and bandwidth can be compared directly across configurations regardless of their absolute signal level. This confirms that camera-based detection, down to a single 10-µm pixel, reproduces the same waveform shape and spectral distribution as conventional EOS. The usable per-pixel bandwidth (~1.5 THz for a single trace) is instead set by the higher per-pixel noise floor, not by spectral distortion of the detection.

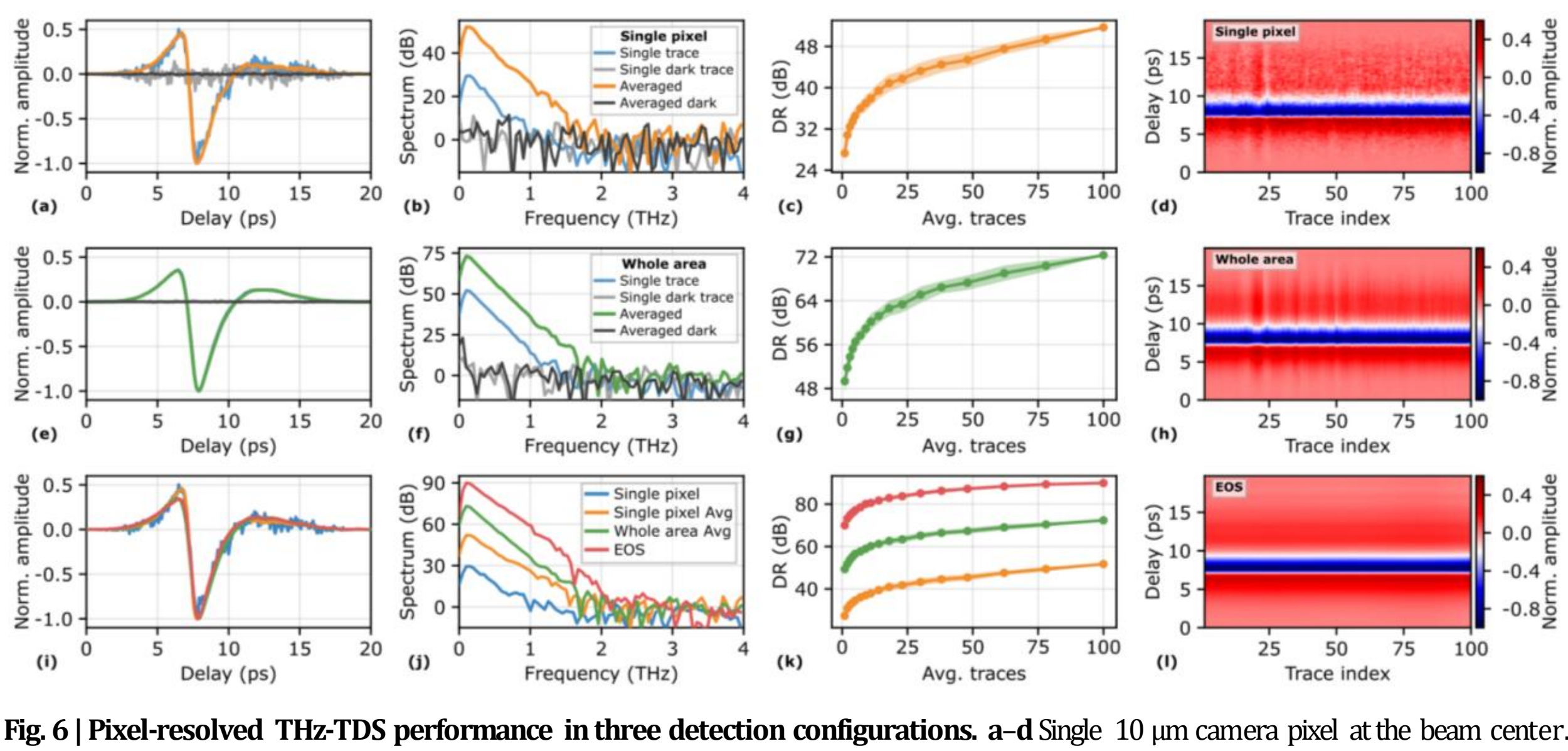

**Fig. 6 | Pixel-resolved THz-TDS performance in three detection configurations. a–d** Single 10 µm camera pixel at the beam center. **e–h** Spatial average over the full probe aperture. **i–l** Balanced-photodetector EOS with the same expanded probe and 5-mm GaP crystal used for camera detection. In each group: **a**, **e**, **i** time-domain waveforms for a single 0.5 s trace and the 100-trace average, with the corresponding dark traces (THz blocked); **b**, **f**, **j** corresponding power spectra, each normalized to its noise floor so that the dark trace lies at 0 dB; **c**, **g**, **k** peak dynamic range (DR) versus number of averaged traces, with the shaded band indicating the standard deviation; **d**, **h**, **l** individual waveforms versus trace index. Peak DR values: single pixel ≈26 dB (single trace) and ≈50 dB (averaged); whole-area average ≈73 dB; EOS ≈90 dB.

**Funding.** Ruhr-Universität Bochum (Open Access Publication Funds); Deutsche Forschungsgemeinschaft (287022738 TRR 196, SFB/TRR 196, for Project S01), RESOLV, THz.NRW, ERC, SERB India under project RJF/2022/000081.

**Acknowledgments**. The research was carried out in the Research Center Chemical Sciences and Sustainability of the University Alliance Ruhr at Ruhr-University Bochum. We thank Francois Blanchard for fruitful discussions on the spatial resolution of our instrument.

**Disclosures.** The authors declare no conflicts of interest.

**Data availability.** The data that support the findings of this study are available from the corresponding author upon reasonable request.

**References**

1. Leitenstorfer, A. *et al.* The 2023 terahertz science and technology roadmap. *J. Phys. D: Appl. Phys.* **56**, 223001 (2023).
2. Guo, X. *et al.* Terahertz nanoscopy: Advances, challenges, and the road ahead. *Applied Physics Reviews* **11**, 021306 (2024).
3. Ulbricht, R., Hendry, E., Shan, J., Heinz, T. F. & Bonn, M. Carrier dynamics in semiconductors studied with time-resolved terahertz spectroscopy. *Rev. Mod. Phys.* **83**, 543–586 (2011).
4. Pezzotti, S. *et al.* Terahertz calorimetry spotlights the role of water in biological processes. *Nat. Rev. Chem.* **9**, 481–494 (2025).
5. Huber, A. J., Keilmann, F., Wittborn, J., Aizpurua, J. & Hillenbrand, R. Terahertz Near-Field Nanoscopy of Mobile Carriers in Single Semiconductor Nanodevices. *Nano Lett.* **8**, 3766–3770 (2008).
6. Cocker, T. L., Jelic, V., Hillenbrand, R. & Hegmann, F. A. Nanoscale terahertz scanning probe microscopy. *Nat. Photon.* **15**, 558–569 (2021).
7. Mitrofanov, O. *et al.* Near-field terahertz probes with room-temperature nanodetectors for subwavelength resolution imaging. *Sci Rep* **7**, 44240 (2017).
8. Chen, A. *et al.* Investigation of water diffusion dynamics in corneal phantoms using terahertz time-domain spectroscopy. *Biomed. Opt. Express* **11**, 1284 (2020).
9. Chen, A. *et al.* Non-contact terahertz spectroscopic measurement of the intraocular pressure through corneal hydration mapping. *Biomed. Opt. Express* **12**, 3438 (2021).
10. Li, X., Li, J., Li, Y., Ozcan, A. & Jarrahi, M. High-throughput terahertz imaging: progress and challenges. *Light Sci Appl* **12**, 233 (2023).
11. Beck, M. *et al.* High-speed THz spectroscopic imaging at ten kilohertz pixel rate with amplitude and phase contrast. *Opt. Express* **27**, 10866 (2019).
12. Chen, S.-C. *et al.* Terahertz wave near-field compressive imaging with a spatial resolution of over λ/100. *Opt. Lett.* **44**, 21 (2019).
13. Stantchev, R. I., Yu, X., Blu, T. & Pickwell-MacPherson, E. Real-time terahertz imaging with a single-pixel detector. *Nat Commun* **11**, 2535 (2020).
14. Li, X. *et al.* Plasmonic photoconductive terahertz focal-plane array with pixel super-resolution. *Nat. Photon.* **18**, 139–148 (2024).
15. Wu, Q., Hewitt, T. D. & Zhang, X.-C. Two-dimensional electro-optic imaging of THz beams. *Appl. Phys. Lett.* **69**, 1026–1028 (1996).
16. Blanchard, F. *et al.* Real-time terahertz near-field microscope. *Opt. Express* **19**, 8277 (2011).
17. Blanchard, F. & Tanaka, K. Improving time and space resolution in electro-optic sampling for near-field terahertz imaging. *Opt. Lett.* **41**, 4645 (2016).
18. Blanchard, F., Arikawa, T. & Tanaka, K. Real-Time Megapixel Electro-Optical Imaging of THz Beams with Probe Power Normalization. *Sensors* **22**, 4482 (2022).
19. Löscher, R. *et al.* Laser-driven few-cycle terahertz sources with high average power. *APL Photon.* **10**, 120901 (2025).

20. Meyer, F., Vogel, T., Ahmed, S. & Saraceno, C. Single-cycle, MHz-repetition rate THz source with 66 mW of average power. *Opt. Lett.* **45**, 2494–2497 (2020).

21. Vogel, T., Mansourzadeh, S. & Saraceno, C. J. Single-cycle, 643 mW average power terahertz source based on tilted pulse front in lithium niobate. *Opt. Lett.* **49**, 4517 (2024).

22. Mansourzadeh, S. *et al.* High-dynamic-range broadband terahertz time-domain spectrometer based on organic crystal MNA. *Photon. Res.* **13**, 2510–2519 (2025).

23. Buldt, J. *et al.* Gas-plasma-based generation of broadband terahertz radiation with 640 mW average power. *Opt. Lett.* **46**, 5256–5259 (2021).

24. Khalili, M., Wang, Y., Winnerl, S. & Saraceno, C. J. High-power single-cycle THz emission from large-area photoconductive emitters at 400 kHz. *Opt. Lett.* **50**, 2141 (2025).

25. Yasui, T. *et al.* Real-time terahertz color scanner for moving objects. *Opt. Express* **16**, 1208 (2008).

26. Dreyhaupt, A., Winnerl, S., Dekorsy, T. & Helm, M. High-intensity terahertz radiation from a microstructured large-area photoconductor. *Appl. Phys. Lett.* **86**, 121114 (2005).

27. Winnerl, S. Scalable Microstructured Photoconductive Terahertz Emitters. *J. Infrared Millim. Terahertz Waves* **33**, 431–454 (2012).

28. Beck, M. *et al.* Impulsive terahertz radiation with high electric fields from an amplifier-driven large-area photoconductive antenna. *Opt. Express* **18**, 9251 (2010).

29. Faure, J., Van Tilborg, J., Kaindl, R. A. & Leemans, W. P. Modelling Laser-Based Table-Top THz Sources: Optical Rectification, Propagation and Electro-Optic Sampling. *Opt. Quantum Electron.* **36**, 681–697 (2004).

30. Garufo, A. *et al.* A Connected Array of Coherent Photoconductive Pulsed Sources to Generate mW Average Power in the Submillimeter Wavelength Band. *IEEE Trans. THz Sci. Technol.* **9**, 221–236 (2019).

31. Huiskes, M. D. *et al.* High THz Average Power Generation from Photoconductive Connected Array Antenna Excited by High-Power Laser. *IEEE Trans. THz Sci. Technol.* 1–11 (2026) doi:10.1109/TTHZ.2026.3669435.

32. Kohlhaas, R. B. *et al.* 637 ***μ***W emitted terahertz power from photoconductive antennas based on rhodium doped InGaAs. *Applied Physics Letters* **117**, 131105 (2020).

33. Vogel, T. *et al.* Performance of Photoconductive Receivers at 1030 nm Excited by High Average Power THz Pulses. *IEEE Trans. THz Sci. Technol.* **14**, 139–151 (2024).

34. Heßelmann, T. *et al.* Ultrafast photoconductors for 1030 nm excitation based on MBE-grown, transition metal-doped InGaAlAs. in *2025 50th international conference on infrared, millimeter, and terahertz waves (IRMMW-THz)* 1–2 (IEEE, 2025).

35. Singh, A. *et al.* Up to 70 THz bandwidth from an implanted Ge photoconductive antenna excited by a femtosecond Er:fibre laser. *Light Sci Appl* **9**, 30 (2020).

36. Cardi, E. *et al.* Ultra-broadband THz emission from one- and two-color laser-induced microplasmas at a high repetition rate. *Optica* **13**, 1294–1302 (2026).

37. Doi, A., Blanchard, F., Tanaka, T. & Tanaka, K. Improving spatial resolution of real-time terahertz near-field microscope. *Journal of Infrared, Millimeter, and Terahertz Waves* **32**, 1043–1051 (2011).

38. Vogel, T. & Saraceno, C. J. Advanced Data Processing of THz-Time Domain Spectroscopy Data with Sinusoidally Moving Delay Lines. *J. Infrared Millim. Terahertz Waves* **45**, 967–983 (2024).

39. Naftaly, M. & Dudley, R. Methodologies for determining the dynamic ranges and signal-to-noise ratios of terahertz time-domain spectrometers. *Opt. Lett.* **34**, 1213 (2009).

40. Scheffter, K. *et al.* Compressed Sensing of Field-Resolved Molecular Fingerprints Beyond the Nyquist Frequency. *Ultrafast Sci* **4**, 0062 (2024).

41. Jiang, Z., Xu, X. G. & Zhang, X.-C. Improvement of terahertz imaging with a dynamic subtraction technique. *Appl. Opt.* **39**, 2982 (2000).

42. Darrow, J. T., Zhang, X.-C., Auston, D. H. & Morse, J. D. Saturation properties of large-aperture photoconducting antennas. *IEEE J. Quantum Electron.* **28**, 1607–1616 (1992).

43. Kim, D. S. & Citrin, D. S. Coulomb and radiation screening in photoconductive terahertz sources. *Applied Physics Letters* **88**, 161117 (2006).

44. Rodriguez, G. & Taylor, A. J. Screening of the bias field in terahertz generation from photoconductors. *Opt. Lett.* **21**, 1046 (1996).

# Supplementary Information

# High-power photoconductive THz emitters for fast parallel electro-optic detection

Mohsen Khalili,[1,*] Robin Löscher [1], Anna Martinez[1] , Abhishek Singh[2], Stephan Winnerl[3], and Clara J. Saraceno[1,4]

[1]Photonics and Ultrafast Laser Science, Ruhr-University Bochum, 44801, Germany
[2]Centre for Advanced Electronics, Indian Institute of Technology Indore, 453552, India
[3]Helmholtz-Zentrum Dresden-Rossendorf, Institute of Ion Beam Physics and Materials Research, 01328 Dresden, Germany
[4]Research Center Chemical Science and Sustainability, University Alliance Ruhr, 44801 Bochum, Germany

*mohsen.khalilikelaki@ruhr-uni-bochum.de

### Supplementary Note 1: Comparison with state-of-the-art THz spectroscopic imaging systems

The performance of the state-of-the-art THz imaging systems discussed in the introduction of the main text is summarized in supplementary Table S1, which lists representative demonstrations of the different approaches together with their key performance parameters: detection method, THz bandwidth, spatial resolution, field of view, acquisition time and imaging throughput.

The figure of merit (FOM) used in Table S1 quantifies hyperspectral imaging throughput. It is defined as the number of spectroscopically resolved pixels acquired per second, FOM = $N_{\text{pixels}}/T_{\text{acq}}$, where $N_{\text{pixels}}$ is the number of spatially resolved pixels and $T_{\text{acq}}$ is the time required to record a complete THz time-domain waveform, and therefore the full amplitude-and-phase spectrum, at every pixel. Because each pixel carries a full waveform, the FOM measures the rate at which spectroscopic information is acquired across the field of view, rather than a raw video frame rate. It thus captures the central difficulty of fast THz time-domain imaging: achieving simultaneously a large number of pixels and a short per-waveform acquisition time. For raster-scanning single-detector systems, $T_{\text{acq}}$ is the per-pixel waveform time multiplied by the number of pixels; for parallel camera- or array-based detectors, $T_{\text{acq}}$ is the time to record one full waveform across all pixels at once, so the FOM scales directly with the number of parallel pixels. The metric is only meaningful for systems that record the full THz waveform per pixel; systems operating at a fixed optical delay or a single frequency (amplitude-only, non-spectroscopic) do not acquire a waveform. Equivalent per-pixel throughput metrics ("pixel rate") have previously been used to benchmark spectroscopic THz imaging systems[1,3].

**Table S1 | Summary of representative state-of-the-art THz spectroscopic imaging systems and their key performance parameters**. The figure of merit (FOM) is defined as the number of spectroscopically resolved pixels acquired per second, FOM = $N_{\text{pixels}}/T_{\text{acq}}$, where $T_{\text{acq}}$ is the time required to acquire a complete THz waveform at every spatial pixel (see accompanying text).

| Reference | Method | THz bandwidth | Resolution | Field of view | $T_{acq}$ | FOM (px/s) |
|---|---|---|---|---|---|---|
| Beck et al[1] | Raster, ASOPS | 3 THz | ~500 µm | cm scale | ~13 s | ~10,000 |
| Blanchard et al.[2] | EO (near-field) | 3 THz | 14 µm (λ/30 @ 0.7 THz) | 370×740 µm² | ~500 s | ~2.8 |
| Yasui et al.[3] | 2D-ST EO line scan | 3 THz | ~mm (diff.-lim.) | 20 mm × 34 ps (1D) | 2 s | 23,200 (1D only) |
| Li et al.[4] | Plasmonic FPA | 3 THz | ~250 µm (~60 µm w/ CNN) | ~4 mm² | ~62 ms | ~1,000 |
| This work | EO camera | ~1.5 THz | ~mm (diff.-lim.) | 14.5 mm² (ø 4.3 mm) | 0.5 s | 289,762 |

EO, electro-optic parallel camera detection; FPA, focal-plane array of photoconductive nanoantennas; 2D-ST, two-dimensional spatiotemporal electro-optic line scanning; diff.-lim., diffraction-limited spatial resolution; CNN, convolutional neural network. $T_{acq}$, acquisition time to record a full THz waveform at every pixel. N/A, not applicable: systems operating at a fixed optical delay or a single frequency do not acquire a per-pixel waveform.

## Supplementary Note 2: Electro-optic response of the GaP detection crystals

In Fig. 2 of the main text, the THz radiation from the large-area photoconductive emitter is characterized by conventional single-pixel electro-optic sampling under the same 295 fs excitation conditions using 3-mm-thick and 0.5-mm-thick GaP detection crystals. Each measured spectrum is normalized to its own peak, and the displayed dark trace belongs to the 3 mm measurement. To interpret the comparison, the incident THz spectrum is multiplied by the complex, frequency-dependent EOS transfer function of the corresponding GaP crystal[5]. Figure S1 shows the two measurements individually, in the time and the frequency domain, together with their dark references and the calculated crystal responses.

The transfer function includes Fresnel transmission at the crystal entrance, the complex THz refractive index and absorption of GaP, phase accumulation caused by mismatch between the THz phase velocity and the optical-probe group velocity, the frequency dependence of the electro-optic coefficient, and temporal averaging imposed by the finite duration and spectrum of the probe pulse. For GaP crystal, the optical-probe spectrum is propagated through the crystal so that group-velocity dispersion before the EOS interaction is also included. The calculation uses the probe-pulse parameters and GaP material model employed to generate the response curves in Fig. 2b of the main text.

The calculated responses remain usable throughout the measured source bandwidth and diverge substantially above approximately 3.5 THz, where the 3 mm crystal consists of pronounced phase-matching structure. In the common reliable range, the measured spectra closely overlap and reach their noise floors at a comparable cutoff. Both traces were averaged over the same measurement time of 156 s, corresponding to 935 individual THz waveforms. Thus, thickness-dependent phase mismatch and absorption in GaP are not the dominant bandwidth limitations over the reported range, which supports a source-limited roll-off. The 3 mm crystal was retained for subsequent measurements because its larger EO signal provides higher SNR.

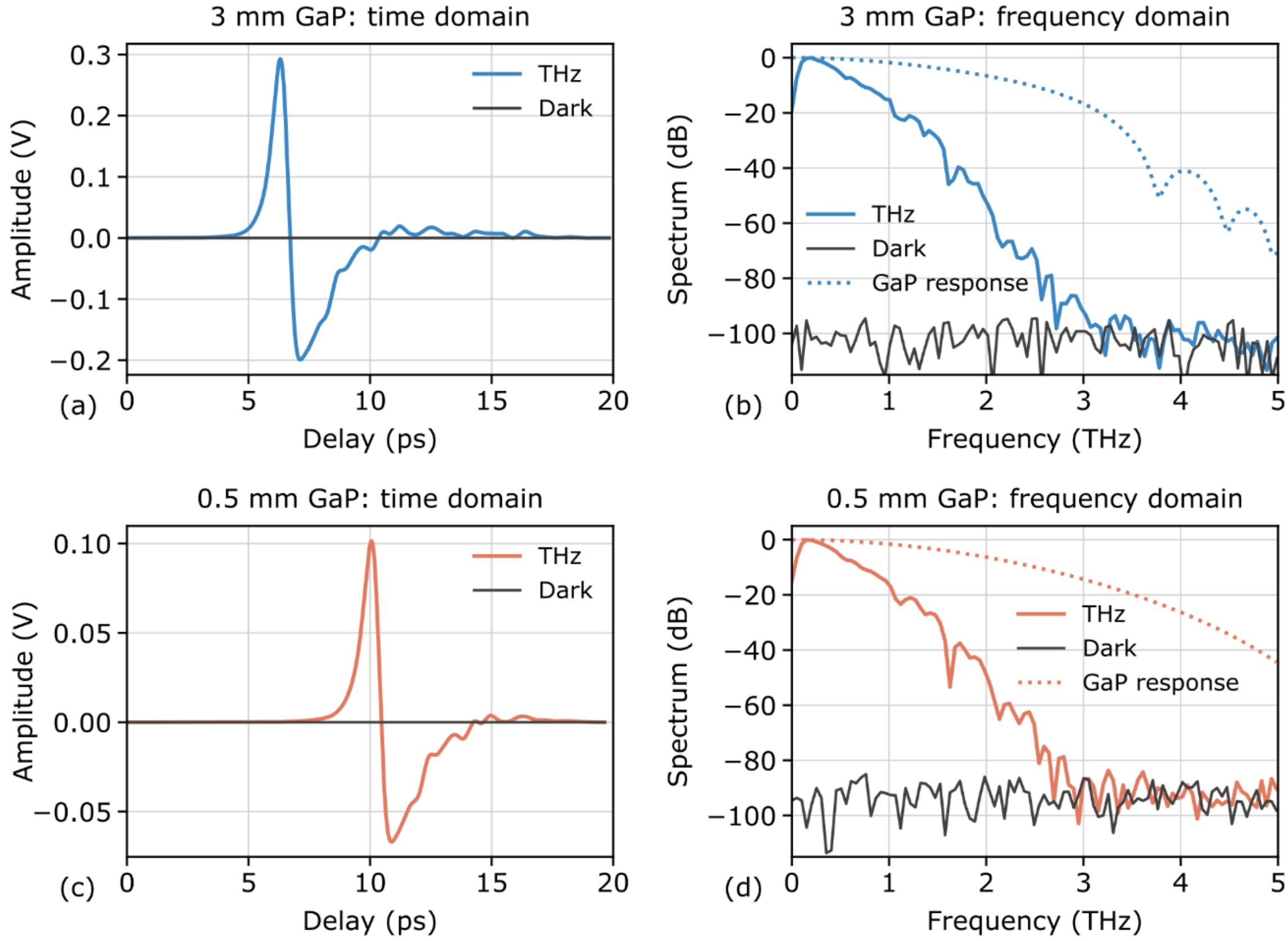

**Fig. S1 | Conventional EOS of the LAE-emitted THz pulse with 3-mm-thick and 0.5-mm-thick GaP detection crystals.** **a,** Time-domain EO signal recorded with the 3 mm crystal, together with the dark trace measured under identical conditions with the bias switched off. **b,** Corresponding power spectra, normalized to the peak of the THz spectrum, with the calculated EOS response of the 3 mm crystal (dotted). **c, d,** The same measurements with the 0.5 mm crystal. The 0.5 mm trace is recorded over a shifted delay window so that the internal reflection of the THz pulse in the thin crystal falls outside the sampled range. The 3 mm crystal gives a peak-to-peak signal of about 0.49 V against 0.17 V for the 0.5 mm crystal, and a dark noise floor roughly 8 dB lower, whereas its calculated response falls off much more steeply above 3 THz.

## Supplementary Note 3: Evolution of the THz Waveform and Spectrum with Pump Power and Bias Voltage

### 1. Pump-Power Dependence at Fixed Bias Voltage

At the lowest bias voltages, increasing the pump power is accompanied by an apparent narrowing of the bipolar THz transient: the negative lobe moves closer to the leading positive lobe and the subsequent recovery becomes shorter. At the same time, the normalized spectral envelope shifts modestly toward higher frequencies and gains relatively high-frequency weight. These changes are most evident in the low-to-intermediate bias range (approximately 1–20 V). At larger bias voltages, the normalized waveforms and spectra progressively converge and become less sensitive to a further increase in pump power. The peak-to-peak EOS amplitude nevertheless increases monotonically with pump power, with increasingly sublinear growth at the upper end of the investigated range.

Pump-fluence-induced THz-pulse narrowing, a shift in the pulse arrival time, and saturation of the peak field and pulse width were previously observed by Hattori et al. in a large-aperture GaAs photoconductive emitter and described using a current-surge model[6]. Chou et al. likewise reported a decreasing pulse width and an increasing spectral peak frequency over a low-to-intermediate fluence range, followed by non-monotonic behavior at higher fluence[7]. These studies provide a qualitative precedent for the evolution observed here; they also show that the trend need not remain monotonic once strong screening, band filling, optical nonlinearities, or heating become important.

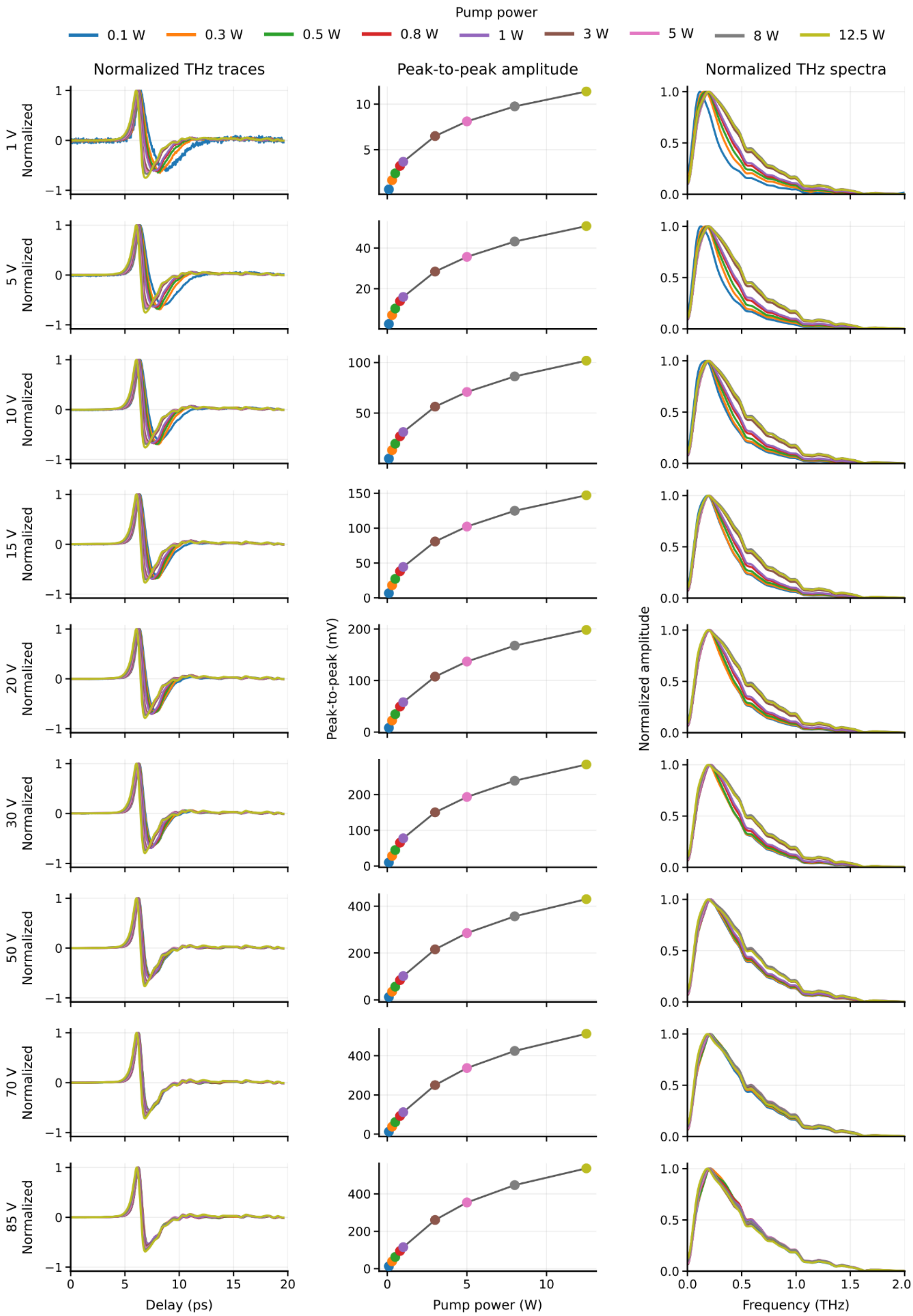

**Fig. S2 | Pump-power dependence of the electro-optically sampled THz emission at fixed bias voltage.** Each row corresponds to one bias voltage (1, 5, 10, 15, 20, 30, 50, 70 and 85 V) and the colors denote pump powers from 0.1 to 12.5 W. Left column: time-domain THz waveforms, each normalized to its own maximum to emphasize changes in pulse shape. Middle column: measured peak-to-peak EOS amplitude as a function of pump power. Right column: corresponding linear-amplitude spectra, each normalized to its own spectral maximum. Because the traces and spectra are normalized individually, absolute signal strengths should be compared using the middle column only.

### 2. Bias-Voltage Dependence at Fixed Pump Power

At low pump power (approximately 0.1–1 W), increasing the bias voltage produces the same qualitative evolution that is seen in Fig. S2 when the pump power is increased at low bias: the THz transient becomes shorter, and the normalized spectrum shifts towards higher frequency. At moderate and high pump powers the normalized temporal and spectral profiles measured at different voltages become closely grouped, so that at simultaneously high pump power and high bias the waveform is comparatively insensitive to a further change of either parameter and the main remaining change is the emitted amplitude. The peak-to-peak amplitude increases monotonically with bias voltage, with a progressively smaller incremental gain towards 70–85 V.

Bias-dependent waveform and spectral changes have also been reported for photoconductive THz emitters. In semi-insulating GaAs, Liu et al. measured a spectral-peak shift from 0.7 to 0.9 THz and bandwidth broadening when the bias was increased from 5 to 20 V; they associated the change with field-driven carrier acceleration and intervalley transport[8]. Three-dimensional carrier-dynamics calculations similarly predict bias-induced pulse shortening and spectral broadening through transfer of electrons from the high-mobility Γ valley to heavier-mass satellite valleys[9]. The behavior is material dependent: Liu et al. observed much less spectral change in their As-implanted GaAs device[8]. Consequently, these studies support the qualitative plausibility of the present trend but do not by themselves establish its microscopic origin in the present emitter.

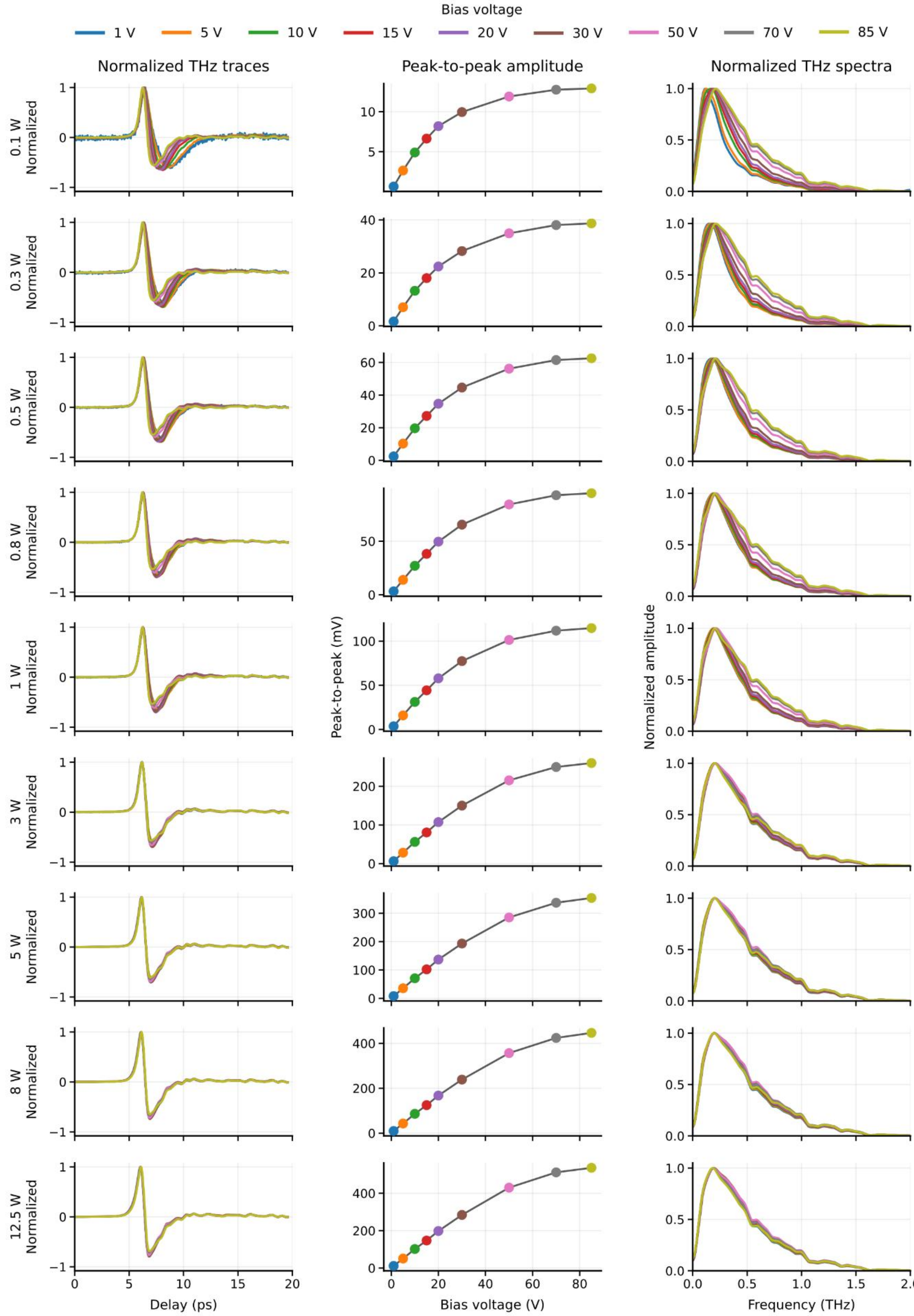

**Fig. S3 | Bias-voltage dependence of the electro-optically sampled THz emission at fixed pump power.** Each row corresponds to one pump power (0.1, 0.3, 0.5, 0.8, 1, 3, 5, 8 and 12.5 W) and the colors denote bias voltages from 1 to 85 V. Left column: independently normalized time-domain THz waveforms. Middle column: measured peak-to-peak EOS amplitude as a function of bias voltage. Right column: linear-amplitude spectra, each normalized to its own maximum. As in Fig. S2, only the middle column carries absolute amplitude information.

To put these observations on a quantitative footing, the spectral peak frequency and the frequency at which the normalized amplitude falls to half of its maximum were extracted from every curve in Figs. S2 and S3. At the lowest bias voltage of 1 V, raising the pump power from 0.1 to 12.5 W shifts the spectral peak from 0.12 to 0.19 THz and the half-maximum frequency from 0.29 to 0.53 THz. Performing the same comparison at 85 V leaves both quantities unchanged. Conversely, at a pump power of 0.1 W, raising the bias from 1 to 85 V shifts the peak from 0.12 to 0.21 THz and the half-maximum frequency from 0.29 to 0.53 THz, whereas at 12.5 W neither quantity changes considerably. Both control parameters therefore drive the emission towards the same limiting spectrum, with a peak near 0.19–0.21 THz and a half-maximum frequency near 0.53 THz, which is reached as soon as either the pump power or the bias is sufficiently large.

The emitted THz field is proportional to the time derivative of the transient photocurrent, so that carrier generation, acceleration, scattering and screening all modify the pulse shape. The observed convergence is consistent with carrier-induced screening of the accelerating field, which is known to alter the waveform and bandwidth of biased photoconductive emitters[10], and at high average power thermal loading can contribute as well[11]. The present data cannot, however, uniquely separate screening, intervalley transport, carrier-lifetime effects and heating.

**Supplementary Note 4: Bias-dependent saturation of the large-area emitter**

Figure 5a of the main text shows the THz average power as a function of optical pump power at three acceleration fields. Here we analyze the three data sets individually, to make the saturation behavior and its dependence on the applied bias explicit. Fig. S4 presents, for each acceleration field, the measured power scaling (a–c), the same data on natural-logarithmic axes where the scaling exponent appears directly as a slope (d–f), and the resulting optical-to-THz conversion efficiency (g–i).

Panels d–f use natural-logarithmic axes because a power law appears there as a straight line. If the emitted power followed $P_{THz} = A\ (P_{opt})^n$, then $\ln(P_{THz}) = \ln(A) + n \ln(P_{opt})$, so that the exponent n is simply the slope of the plotted data. Because the logarithm of a dimensional quantity is not defined, the axes show the logarithm of the numerical value of each quantity expressed in the stated unit, that is $\ln(P_{opt} / 1\ W)$ on the x-axis and $\ln(P_{THz} / 1\ mW)$ on the y-axis. A different choice of unit only shifts the data along the axes by a constant and leaves the slope, which is the quantity of interest, unchanged.

In the absence of screening, the radiated THz field is proportional to the photo-induced current and therefore to the photogenerated carrier density, so the THz average power is expected to scale quadratically with optical pump power. Panels a–c show the measured data together with this quadratic dependence, anchored to the two lowest pump powers of each data set. At the lowest powers, the local logarithmic slopes $d \ln(P_{THz})/d \ln(P_{Opt})$ are 1.88, 1.98 and 1.88 for 20, 51 and 75 kV/cm, respectively, close to the unsaturated value of 2. With increasing pump power, the data fall progressively below the quadratic line; at 12.6 W the local slopes are 0.59, 0.73 and 0.87. Straight-line fits above 6 W give exponents of 0.82, 1.04 and 1.12, confirming that the high-power roll-off is strongest at the lowest bias.

Because the measured curves are not described by a single power law, the exponent decreases as the pump power increases, the exponent is not obtained from one straight-line fit, but evaluated locally at every measured point from its two neighbors, $n(P_i) = [\ln P_{THz}(P_{i+1}) - \ln P_{THz}(P_{i-1})] / [\ln P_{opt}(P_{i+1}) - \ln P_{opt}(P_{i-1})]$. This local logarithmic slope is the exponent of the power law that describes the data in the immediate vicinity of $P_i$. At the first and the last point of each data set only one neighbor is available, and the corresponding two-point difference is used. As an example, the value quoted above for the lowest pump power at 20 kV/cm follows directly from the two lowest measurements of that data set, $P_{opt} = 0.3$ W with $P_{THz} = 0.0363$ mW and $P_{opt} = 0.5$ W with $P_{THz} = 0.0946$ mW, giving $n = [\ln(0.0946) - \ln(0.0363)] / [\ln(0.5) - \ln(0.3)] = 0.958 / 0.511 = 1.88$. Because the values at the two ends of each data set rest on only two measurements, they are less precise than the interior ones: propagating the standard deviations of the individual power measurements gives an uncertainty of approximately ±0.2 for the exponents at the lowest pump power and ±0.1 for those at 12.6 W, whereas the straight-line fits above 6 W, which use seven points each, are precise to approximately ±0.02.

The operational onset of the rollover shifts to higher pump power as the acceleration field increases. Defining the onset by a local slope of 1.5 gives 0.9, 1.8 and 2.4 W for 20, 51 and 75 kV/cm, corresponding to peak on-axis pump fluences of approximately 12, 23 and 31 μJ/cm² for a 7.0 mm $1/e^2$ pump diameter at 400 kHz. The rollover is consistent with carrier-induced screening of the effective accelerating field. Both the carrier-separation field and the radiated THz near field oppose the applied bias, and studies of large-aperture GaAs photoconductors indicate that prompt near-field screening can dominate during THz generation[12,13]. The weaker roll-off at higher bias suggests that the stronger applied field partly compensates the carrier-induced field reduction or modifies carrier extraction, but the bias dependence alone does not identify a unique microscopic mechanism. Optical heating, carrier scattering and current-related limitations may also contribute. Related bias-dependent saturation has been reported for a large-area GaAs emitter operated at lower fields and repetition rate[14].

The dotted vertical line in every panel of Fig. S4 marks the pump power at which this local slope has fallen to 1.5, that is midway between the unsaturated value of 2 and purely linear scaling. This threshold is a convention adopted here so that a single comparable number can be quoted for each data set; it is not a physical constant, and the absolute onset powers depend on the value chosen. The comparison between the three acceleration fields does not depend on that choice, as Table S2 shows: whichever threshold between 1.8 and 1.2 is used, the onset always occurs at the lowest pump power for 20 kV/cm and at the highest for 75 kV/cm, and the onset powers relative to the 20 kV/cm value stay close to 2 for 51 kV/cm and to 3 for 75 kV/cm.

**Table S2 | Sensitivity of the roll-off onset to the chosen slope threshold.** Pump power at which the local logarithmic slope $d \ln P_{THz} / d \ln P_{opt}$ falls to the value in the first column, for each of the three acceleration fields, together with the same onset powers normalized to the 20 kV/cm value. The absolute onset powers depend strongly on the threshold; their ordering and their ratios do not.

| Slope threshold | 20 kV/cm | 51 kV/cm | 75 kV/cm | ratio 51 / 20 | ratio 75 / 20 |
|---|---|---|---|---|---|
| 1.8 | 0.49 W | 0.73 W | 1.12 W | 1.5 | 2.3 |
| 1.7 | 0.61 W | 1.06 W | 1.52 W | 1.7 | 2.5 |
| 1.5 | 0.90 W | 1.82 W | 2.43 W | 2.0 | 2.7 |
| 1.4 | 1.08 W | 2.30 W | 3.13 W | 2.1 | 2.9 |
| 1.2 | 1.79 W | 3.87 W | 5.45 W | 2.2 | 3.1 |

The practical consequence is shown in panels g–i. At 20 kV/cm the incident optical-to-THz conversion efficiency reaches approximately 0.038% near 4 W and then decreases by about 15% up to 12.6 W. At 51 kV/cm the maximum of approximately 0.147% occurs near 10 W. At 75 kV/cm the efficiency remains statistically near its maximum through 12.6 W: it reaches 0.237% at 12 W and 0.235% at 12.6 W. A larger bias therefore increases the THz output and delays the pronounced high-power roll-off, but the sub-quadratic power scaling shows that the emitter is not strictly unsaturated. Further average-power scaling will require balancing carrier screening and transport with optical and electrical thermal loading.

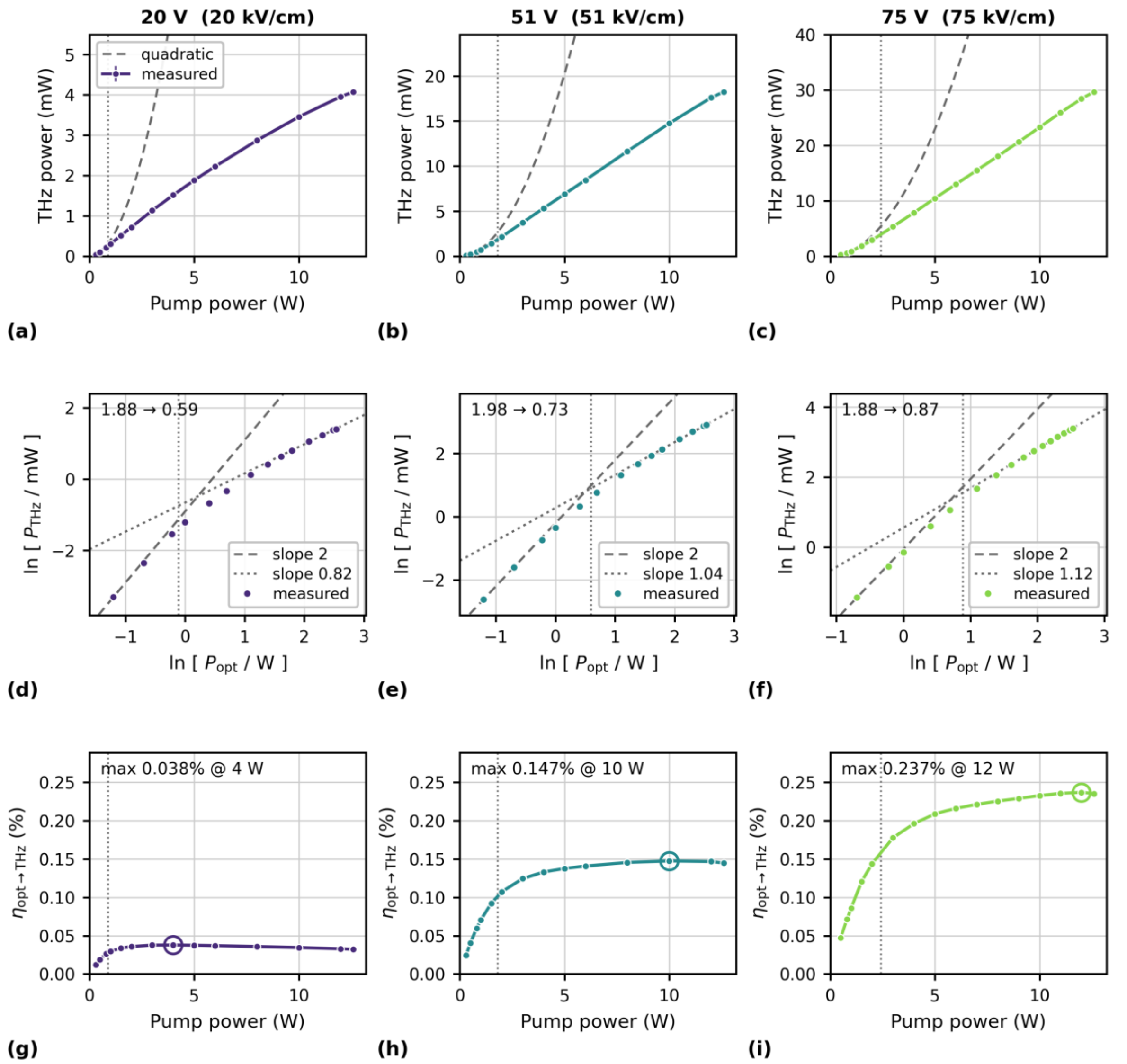

Fig. S4 | Bias-dependent saturation of the large-area photoconductive emitter. Columns correspond to acceleration fields of 20, 51 and 75 kV/cm (bias voltages of 20, 51 and 75 V); the colours match Fig. 5a of the main text. a–c, Measured THz average power versus optical pump power (symbols, with error bars) together with the quadratic dependence expected in the absence of saturation (dashed), anchored to the two lowest pump powers. d–f, The same data on natural-logarithmic axes, where the scaling exponent is the slope. The dashed line has slope 2, the unsaturated limit; the dotted line is a straight-line fit for pump powers above 6 W, whose slope is given in the legend. The pair of values in the upper left gives the local slope at the lowest and at the highest pump power. g–i, Incident optical-to-THz conversion efficiency versus optical pump power, with the maximum marked by an open circle. The dotted vertical line in every panel marks the pump power at which the local slope has fallen to 1.5. The dashed line in a–c is that quadratic law, i.e. the behaviour expected if no saturation occurred. The threshold of 1.5 is a convention adopted so that the three data sets can be compared by a single number; it is not a physical constant. The local logarithmic slope, the logarithmic axes and the choice of threshold are defined in the accompanying text.

### Supplementary Note 5: Repetition-rate dependence and pulse-energy scaling

To separate the influence of the pump pulse energy from that of the repetition rate on the emitter performance, the laser repetition rate was swept from 50 kHz to 400 kHz with the pump spot diameter fixed at 7 mm and the acceleration field fixed at 55 kV/cm, the PWM bias being synchronized to the corresponding repetition rate in each case. Each repetition rate then covers a different window of pump pulse energy: 1.25–6.25 µJ at 400 kHz (average power = 0.5-2.5 W), rising to 10–30 µJ at 50 kHz (average power = 0.5-1.5 W), corresponding to peak on-axis fluences of 6–32 and 52–156 µJ/cm² for the 7 mm spot. Plotting the THz pulse energy against the pump pulse energy, rather than comparing average powers, therefore probes the emitter over more than one order of magnitude of excitation fluence.

In the unsaturated limit the radiated THz field is proportional to the photo-induced current and hence to the photogenerated carrier density, so the THz pulse energy is expected to scale with the square of the pump pulse energy. Figure S5a shows that this expectation holds at low pulse energy and is progressively lost as the pulse energy grows: the data follow the quadratic reference up to roughly 5 µJ and fall increasingly below it thereafter. Fitting a power law $E_{THz} \propto E_{Opt}^{n}$ separately to each repetition rate returns the exponents listed in Table S3, which fall monotonically from 1.82 at 400 kHz to 1.07 at 50 kHz. The nearly linear behavior of the low-repetition-rate curves is therefore not the intrinsic response of the emitter but the saturated tail of the same quadratic law, reached because those measurements sit at the highest pulse energies.

**Table S3 | Power-law exponent of the THz pulse energy for each repetition rate.** Exponent n obtained from a straight-line fit of ln $E_{THz}$ against ln $E_{opt}$ over the pulse-energy range accessible at that repetition rate, together with that range and the corresponding peak on-axis fluence for the 7 mm pump spot. The exponent is a property of the fluence window covered, not of the repetition rate itself: see Fig. S5b.

| Repetition rate | Pump pulse energy (µJ) | Peak fluence (µJ/cm²) | Exponent n |
|---|---|---|---|
| 400 kHz | 1.25 – 6.25 | 6 – 32 | 1.82 |
| 200 kHz | 2.50 – 12.5 | 13 – 65 | 1.65 |
| 133 kHz | 3.76 – 18.8 | 20 – 98 | 1.52 |
| 100 kHz | 5.00 – 25.0 | 26 – 130 | 1.38 |
| 80 kHz | 6.25 – 30.0 | 32 – 156 | 1.18 |
| 66 kHz | 7.58 – 30.3 | 39 – 157 | 1.13 |
| 50 kHz | 10.0 – 30.0 | 52 – 156 | 1.07 |

That the roll-off is governed by pulse energy alone, and not by the repetition rate, is shown directly in Fig. S5b, where the local exponent d ln $E_{THz}$ / d ln $E_{opt}$ is evaluated point by point. All seven repetition rates fall on a single curve: the exponent decreases from 1.88 at 1.25 µJ, through 1.5 near 8 µJ, to 0.7–0.8 at 30 µJ. Measurements taken at the same pulse energy, but different repetition rates return the same exponent to within a few per cent; at 5 µJ, for example, it is 1.66, 1.70 and 1.72 at 100, 200 and 400 kHz, respectively. The roll-off is thus set by the excitation fluence, and therefore by the carrier-induced screening discussed in Supplementary Note 4.

The repetition rate itself has a further, much smaller influence, which becomes visible when measurements with similar pump pulse energy are compared. At a fixed pulse energy the fluence is unchanged, so the difference must come from the average power or heat load rather than from screening. Raising the repetition rate from 50 to 200 kHz at a pulse energy of 10 µJ, which raises the average power from 0.5 to 2.0 W, decreases the THz pulse energy from 15.34 to 12.88 nJ, a loss of 16%. The same fourfold step from 100 to 400 kHz at 5 µJ, again from 0.5 to 2.0 W, costs 13% (4.43 to 3.87 nJ). This residual effect is much smaller than the fluence-driven roll-off described above, but it is systematic, and it is why the efficiency curves in Fig. S5c do not lie exactly on top of one another. The conversion efficiency in Fig. S5c rises steeply with pulse energy, peaks at approximately 0.18% near 20 µJ and then declines, mirroring the exponent falling below one in Fig. S5b.

Two practical conclusions follow. First, at a fixed average power the emitter is more efficient at a lower repetition rate, because the same energy is then delivered in fewer but stronger pulses. At an average power of 1 W the efficiency is 0.18% at 50 kHz, where each pulse carries 20 µJ, but only 0.045% at 400 kHz, where each pulse carries 2.5 µJ. This advantage persists until the fluence-driven saturation sets in above about 20 µJ. Second, saturation is governed by the fluence rather than by the repetition rate, so it can be postponed by enlarging the excited area instead of by changing the pulse timing. A large-area emitter can therefore be used in either regime: at a high repetition rate it provides high average power, while at a low repetition rate it provides high pulse energy, and hence high peak fields.

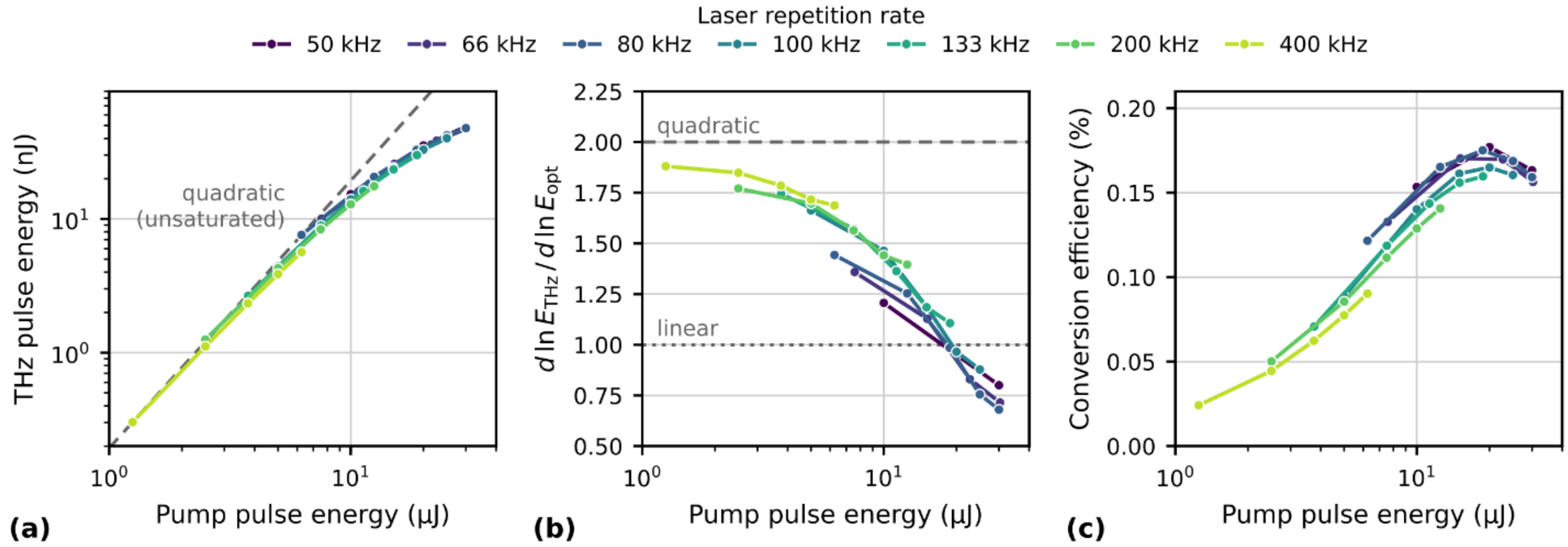

**Fig. S5 | Pulse-energy scaling of the large-area emitter across repetition rates.** Measurements at seven laser repetition rates from 50 to 400 kHz, at a fixed acceleration field of 55 kV/cm and a 7 mm pump spot; the colors match Fig. 5c of the main text. One microjoule of pump pulse energy corresponds to a peak on-axis fluence of 5.2 μJ/cm². **a,** THz pulse energy versus pump pulse energy on logarithmic axes. The dashed line is the quadratic dependence expected in the absence of saturation, anchored to the lowest measured point. **b,** Local exponent $d \ln E_{THz} / d \ln E_{opt}$, evaluated point by point. All repetition rates collapse onto one curve, showing that the scaling is governed by the pump pulse energy alone. **c,** Optical-to-THz conversion efficiency versus pump pulse energy. At a given pulse energy the efficiency is slightly lower for the higher repetition rate, because the emitter then carries a larger average power.

## Supplementary Note 6: Single-pixel waveform characterization across the field of view

To verify that spectroscopic information is available at the level of individual camera pixels, THz waveforms were extracted at five representative single-pixel positions of the registered EO image (center, up, down, left, right; marked in Fig. 3c of the main text and Fig. S6h), and compared with the spatial mean over the whole illuminated area. For each position, N = 100 consecutive THz traces were recorded, together with the corresponding dark traces (N = 101) acquired with the bias voltage switched off, which serve as the noise reference.

Figure S6 summarizes the results. The first column shows the time-domain waveforms: a representative single trace (selected as the trace with the maximum peak-to-peak amplitude), the corresponding single dark trace, and the final averages of THz and dark traces. Even without averaging, the single-pixel, single-trace waveforms clearly resolve the THz transient, and the average single-pixel waveforms closely reproduce the pulse shape of the whole-area mean, confirming uniform temporal response across the field of view.

The second column shows the corresponding power spectra. Each spectrum is normalized to the noise floor of its corresponding dark trace (single or averaged), within the detected bandwidth. The peak dynamic range ($DR_{peak}$) and the bandwidth over which the spectrum exceeds the noise floor by more than 10 dB (dashed lines) are indicated for each position.

The whole-area mean reaches $DR_{peak}$ = 73.1 dB with a 10 dB bandwidth of 1.56 THz, while the individual pixels reach $DR_{peak}$ = 43.5–52.0 dB with 10 dB bandwidths of 1.00–1.56 THz for the same 100-trace average. The difference between the whole-area and single-pixel DR primarily reflects the spatial averaging over ~$10^5$ pixels, which suppresses the per-pixel noise, whereas the single-pixel values represent the true per-pixel sensitivity of the camera-based detection.

The third column shows all 100 single traces per position as a heatmap of EO signal versus delay and trace index, demonstrating the shot-to-shot stability of the waveform at individual pixels over the full acquisition: the arrival time and polarity of the main transient are stable over all traces, with no drift of the delay axis. The bottom row overlays the averaged waveforms and spectra of all positions.

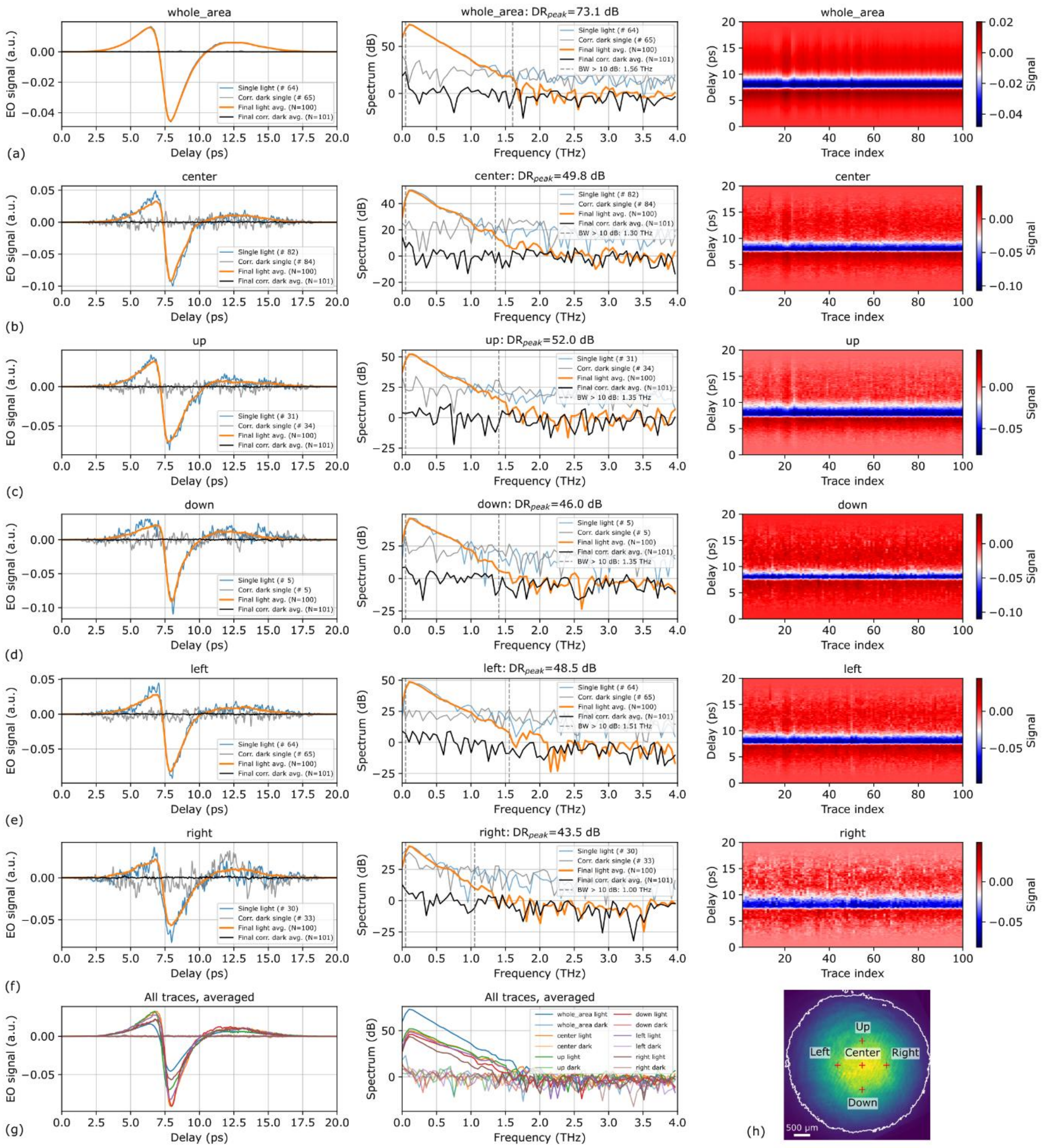

**Fig. S6 | Single-pixel THz waveform and spectrum characterization.** Rows: whole-area mean and five single-pixel positions (center, up, down, left, right) marked in h. Left column: time-domain EO signal, showing a representative single light trace, the corresponding single dark trace, and the final averages over N = 100 light and N = 101 dark traces. Middle column: power spectra of the same signals, normalized to the mean noise floor of the averaged dark spectrum within the detected band, such that the dark reference lies at 0 dB; DR_peak denotes the peak dynamic range and the dashed lines mark the band where the spectrum exceeds the noise floor by >10 dB. Right column: heatmap of all 100 consecutive single light traces (EO signal versus delay and trace index), illustrating shot-to-shot waveform stability. Bottom row: overlay of the final average waveforms and spectra of all positions. h, Registered EO image of the THz beam with the five single-pixel positions marked.

**Supplementary Note 7: Optical characterization of the fundamental and second-harmonic pulses**

Figure S7 shows the optical spectra and intensity autocorrelations of the TruMicro fundamental and its second harmonic. The 1030 nm fundamental provides both the input for second-harmonic generation and the electro-optic probe, while the green pulses excite the large-area emitter. Gaussian fits to the measured spectra give center wavelengths of 1030.9 nm and 515.5 nm, with spectral intensity FWHM bandwidths of 8.1 nm and 2.0 nm, respectively.

Each spectrum and autocorrelation was fitted independently by unweighted least squares to a Gaussian with free amplitude, center, FWHM and constant background, using the complete recorded range. Fits used the exported delay coordinates, spanning approximately 16.7 ps for the fundamental and 15 ps for the second harmonic. Their fitted intensity-autocorrelation FWHM values are 417 fs and 358 fs. Dividing these widths by √2, as appropriate for Gaussian intensity pulses, gives pulse durations of 295 fs and 253 fs, respectively.

The fundamental autocorrelation exhibits a weak broad pedestal that was observed both with and without acquisition averaging. The longer delay range samples the background farther from this pedestal. The reported duration is therefore a full-range Gaussian-equivalent intensity FWHM. Similarly, the quoted spectral bandwidths are Gaussian-fit estimates of spectra that are not exactly Gaussian.

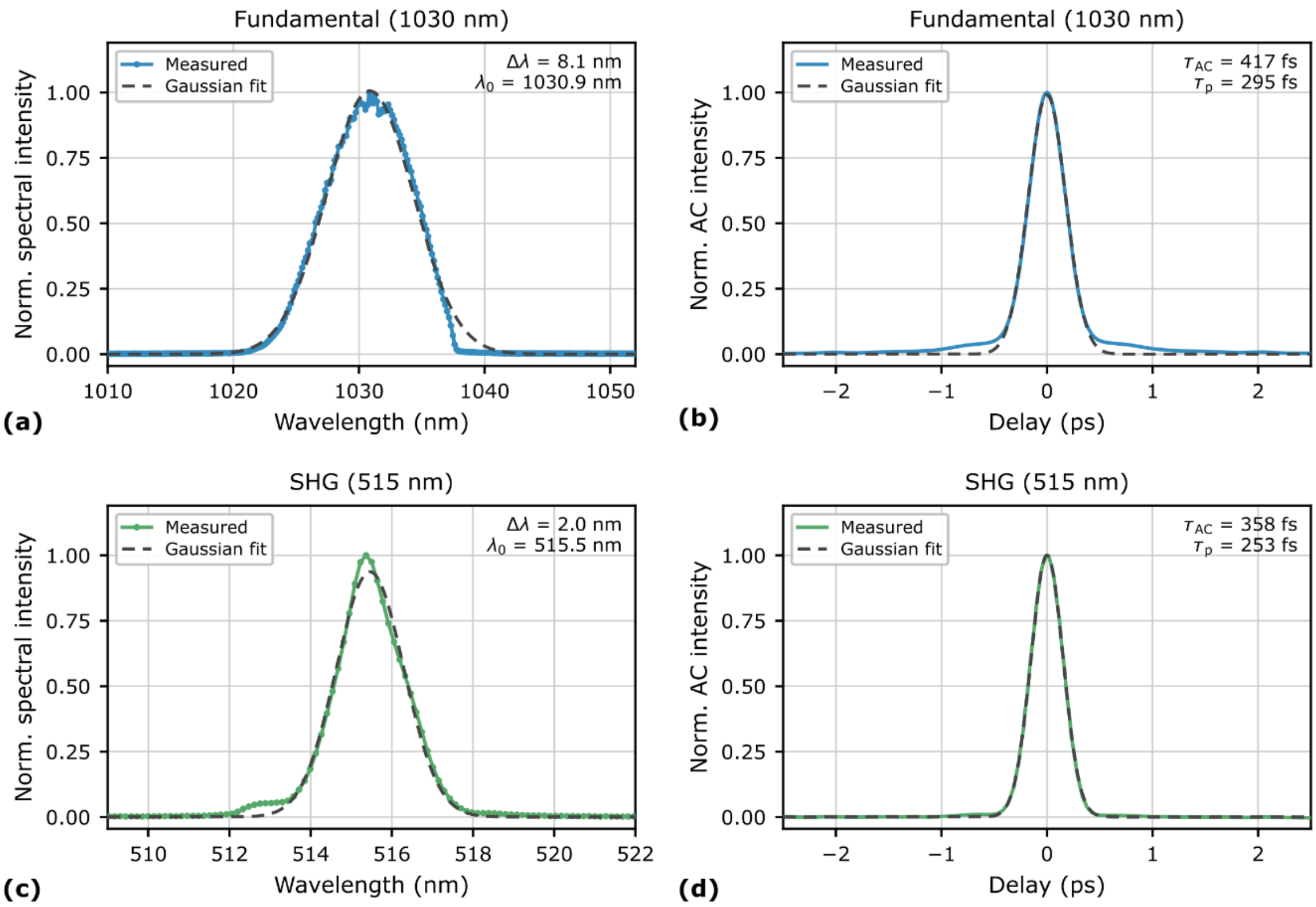

**Fig. S7 | Optical spectra and intensity autocorrelations of the fundamental and second-harmonic pulses.** a, b, Fundamental spectrum and intensity autocorrelation. c, d, Corresponding measurements for the green second harmonic. Solid curves show the measurements; dashed curves show Gaussian fits. Signals are normalized to the measured peak after subtraction of the fitted constant background. Autocorrelation delays are centered at the fitted peak, and only the central regions are displayed; all fits use the complete recorded ranges. The pulse duration is the autocorrelation FWHM divided by √2.

**Supplementary References**

1. Beck, M. *et al.* High-speed THz spectroscopic imaging at ten kilohertz pixel rate with amplitude and phase contrast. *Opt. Express* **27**, 10866 (2019).
2. Blanchard, F. *et al.* Real-time terahertz near-field microscope. *Opt. Express* **19**, 8277 (2011).
3. Yasui, T. *et al.* Real-time terahertz color scanner for moving objects. *Opt. Express* **16**, 1208 (2008).
4. Li, X. *et al.* Plasmonic photoconductive terahertz focal-plane array with pixel super-resolution. *Nat. Photon.* **18**, 139–148 (2024).
5. Faure, J., Van Tilborg, J., Kaindl, R. A. & Leemans, W. P. Modelling Laser-Based Table-Top THz Sources: Optical Rectification, Propagation and Electro-Optic Sampling. *Opt. Quantum Electron.* **36**, 681–697 (2004).
6. Toshiaki Hattori, T. H., Keiji Tukamoto, K. T. & Hiroki Nakatsuka, H. N. Time-Resolved Study of Intense Terahertz Pulses Generated by a Large-Aperture Photoconductive Antenna. *Jpn. J. Appl. Phys.* **40**, 4907 (2001).
7. Chou, R.-H., Liu, T.-A. & Pan, C.-L. Analysis of terahertz pulses from large-aperture biased semi-insulating and arsenic-ion-implanted GaAs antennas. *Journal of Applied Physics* **104**, 053121 (2008).
8. Liu, T.-A., Tani, M. & Pan, C.-L. THz radiation emission properties of multienergy arsenic-ion-implanted GaAs and semi-insulating GaAs based photoconductive antennas. *Journal of Applied Physics* **93**, 2996–3001 (2003).
9. Castro-Camus, E., Lloyd-Hughes, J. & Johnston, M. B. Three-dimensional carrier-dynamics simulation of terahertz emission from photoconductive switches. *Phys. Rev. B* **71**, 195301 (2005).
10. Pedersen, J. E. *et al.* Ultrafast local field dynamics in photoconductive THz antennas. *Applied Physics Letters* **62**, 1265–1267 (1993).
11. Khalili, M. *et al.* Microstructured large-area photoconductive terahertz emitters driven at high average power. *Opt. Express* **32**, 22955–22969 (2024).
12. Darrow, J. T., Zhang, X.-C., Auston, D. H. & Morse, J. D. Saturation properties of large-aperture photoconducting antennas. *IEEE J. Quantum Electron.* **28**, 1607–1616 (1992).

13. Rodriguez, G. & Taylor, A. J. Screening of the bias field in terahertz generation from photoconductors. *Opt. Lett.* **21**, 1046 (1996).

14. Beck, M. *et al.* Impulsive terahertz radiation with high electric fields from an amplifier-driven large-area photoconductive antenna. *Opt. Express* **18**, 9251 (2010).